\documentclass[conference]{IEEEtran}
\IEEEoverridecommandlockouts                             

\usepackage{tabularx} 
\usepackage{amsmath}  
\usepackage{amssymb}  
\usepackage{amsfonts}
\usepackage{graphicx} 
\usepackage{cite} 
\usepackage{enumitem}  
\usepackage[final]{hyperref} 
\usepackage{steinmetz} 
\usepackage{mathrsfs}  
\usepackage{multirow} 
\usepackage{stfloats}
\usepackage{float}
\usepackage{mathptmx} 
\usepackage{anyfontsize}
\usepackage{t1enc}
\usepackage{diagbox} 
\usepackage{caption} 
\usepackage{subcaption} 

\usepackage{xcolor}
\usepackage{lipsum}
\newcommand*{\mybox}[2]{\colorbox{#1!10}{\parbox{.98\linewidth}{#2}}}

\usepackage{multirow} 
\def\BibTeX{{\rm B\kern-.05em{\sc i\kern-.025em b}\kern-.08em
    T\kern-.1667em\lower.7ex\hbox{E}\kern-.125emX}}

\hypersetup{
	colorlinks=true,       
	linkcolor=black,        
	citecolor=black,        
	filecolor=magenta,     
	urlcolor=black         
}

\begin{document}

\title{Hierarchical Frequency and Voltage Control using Prioritized Utilization of Inverter Based Resources}

\author{Rahul Chakraborty$^{1}$, \textit{Student Member, IEEE}, Aranya Chakrabortty$^{1}$, \textit{Senior Member, IEEE},\\ Evangelos Farantatos$^{2}$,  \textit{Senior Member, IEEE}, Mahendra Patel$^{2}$, \textit{Fellow, IEEE},\\ Hossein Hooshyar$^{3}$, \textit{Senior Member, IEEE}, and Atena Darvishi$^{3}$, \textit{Member, IEEE}
\thanks{$^{1}$R. Chakraborty and A. Chakrabortty are with the Department of Electrical and Computer Engineering,
        North Carolina State University, USA.
        {(e-mail:\tt\small \{rchakra5, achakra2\}@ncsu.edu)}}%
\thanks{$^{2}$E. Farantatos and M. Patel are with Electric Power Research Institute (EPRI), USA.
        {(e-mail: \tt\small \{efarantatos, mpatel\}@epri.com)}}%
\thanks{$^{3}$H. Hooshyar and A. Darvishi are with New York Power Authority (NYPA), USA.
        {(e-mail: \tt\small \{Hossein.Hooshyar, Atena.Darvishi\}@nypa.gov)}}%
}

\maketitle
\thispagestyle{plain}  
\pagestyle{plain}      

\begin{abstract}
We propose a novel hierarchical frequency and voltage control design for multi-area power system integrated with inverter-based resources (IBRs). The design is based on the idea of prioritizing the use of IBRs over conventional generator-based control in compensating for sudden and unpredicted changes in loads and generations, and thereby mitigate any undesired dynamics in the frequency or the voltage by exploiting their fast actuation time constants. A new sequential optimization problem, referred to as Area Prioritized Power Flow (APPF), is formulated to model this prioritization. It is shown that compared to conventional power flow APPF not only leads to a fairer balance between the dispatch of active and reactive power from the IBRs and the synchronous generators, but also limits the impact of any contingency from spreading out beyond its respective control area, thereby guaranteeing a better collective dynamic performance of the grid. This improvement, however, comes at the cost of adding an extra layer of communication needed for executing APPF in a hierarchical way. Results are validated using simulations of a 9-machine, 6-IBR, 33-bus, 3-area power system model, illustrating how APPF can mitigate a disturbance faster and more efficiently by prioritizing the use of local area-resources.

\textit{Index Terms} - hierarchical control, frequency control, voltage control, renewable energy, inverter-based resources.
\end{abstract}

\section{Introduction}\label{introduction}
Conventional methods used for both frequency control and voltage control in today's electric transmission grid need to be revisited over the coming decade as more  power electronic inverter-based resources (IBRs) such as wind, solar and energy storage devices penetrate power systems across the world. IBRs have significantly faster time-constants in following their active and reactive power dispatch commands compared to droop control and automatic generation control (AGC) \cite{2018_Milano_Dorfler_Hill,2017_B_Johnson_100_IBR}. Therefore, it will only be natural for grid operators to {\it prioritize} the use of IBRs over synchronous generators (SG) for frequency and voltage control not only for steady-state regulation but also for faster mitigation of unwanted dynamics. With millions of IBRs anticipated to penetrate power systems over the foreseeable future, limiting these control methods to how they are executed currently will be a tremendous under-utilization of renewables in terms of how they typically cope with the grid dynamics. Of course, given the significant inertial effect of SGs, one cannot expect the renewables to carry out these control efforts solely by themselves \cite{2013_Vijay_Vittal_Wind,2014_Andreas_Goran_Anderson,2018_Zhao_Dhople}, but with a right combination of engaging both SGs and IBRs, with priority given to IBRs whenever possible, one may be able to improve control efficiency by a notable extent. 

Quantification of such prioritization, however, has been addressed very sparsely in the power system control literature so far. Several seminal papers have been written on how IBRs can facilitate frequency regulation using mathematical tools from both linear and nonlinear control theory such as phase cohesiveness \cite{2017_Garcia_freq_reg}, passivity \cite{2019_Ali_Rafael_Chakrabortty,2020_Awal_Yu_TPEL}, synchronization in complex oscillator networks \cite{2013_Dorfler_Chertkov}, sliding mode control \cite{2017_Su}, model predictive control \cite{labella2020supervised}, and event-triggered $\ell_{\infty}$-control \cite{2020_Yang_Liu_Hill}, to name a few, but not related to prioritization over synchronous generation. The same argument applies to the literature on voltage control \cite{2016_Mani_Voltage,2017_L_Mili,2018_Hernandez}. Several innovative IBR control methods have been proposed in the context of microgrids such as in \cite{2011_Gurrero_ac_dc,2015_Porco_Dorfler_Bullo_Freq_reg_paper, 2017_Guerrero_Hierarchical_Control}, but their application in high-voltage transmission grids in the presence of conventional generation controls has not been addressed.  

Motivated by this problem, in this paper we propose a new control methodology by which the use of IBRs can be prioritized for both frequency and voltage control. The fundamental concept of the proposed frequency control is as follows. We consider a power system divided into a finite set of distinct and non-overlapping control areas. The definition of these areas is open to the operator, and can be, for instance, the same as the balancing regions that are used for AGC or smaller regional areas. Each area should have IBRs which satisfy a certain percentage of the total generation specified by the operator. When a sudden unpredicted change in generation or load occurs in any area, we first employ primary control based on the available headroom of the renewable resources in that area. This contingent area is tagged as the so-called first {\it hierarchy}, while areas that share direct tie-line connections with the contingent area are collectively referred to as the second hierarchy. The areas that share tie-line connections with the second hierarchy are collectively referred to as the third hierarchy, and so on. In parallel to fast re-dispatch of available IBR headroom in primary control, we run a new setpoint dispatch method, referred to as Area-Prioritized Power Flow (APPF) by which the active and reactive power setpoints for the IBRs across the different hierarchies can be computed to bring the system-wide frequency back to synchronous while prioritizing the availability of the renewables in each hierarchy, starting from the contingent area or the first hierarchy itself. The optimization is done in multiple steps in a sequential way from one hierarchy to the next. First, the optimal active power setpoints of the IBRs in the contingent area are computed by formulating the APPF as a power balancing problem that minimizes the disruption of the flow in the tie-lines connecting the contingent area to the second hierarchy. If the available IBR capacity in the contingent area is less than the power imbalance, then the deficit amount of power is provided from the IBRs in the second hierarchy with the same objective of minimal disruption in the tie-lines. This step ensures that the power balance is satisfied locally inside the contingent area with minimal disruption of dynamics in the rest of the system. This step alone, however, does not bring back the frequency to the synchronous value as power is only balanced locally in the contingent area. Therefore, a second step of APPF is carried out to recompute the IBR setpoints in the second hierarchy to balance out the deficit power while making sure that the tie-line flows it shares with its next hierarchy is minimally interrupted. If the total IBR capacity in the second hierarchy is not sufficient for the balance, then a third step of APPF is applied to the IBRs in the third hierarchy, and so on. The process is continued until all power flows are balanced, and the system frequency is back to synchronous. The synchronous generators are allowed to run AGC as usual in the backdrop while the different steps of APPF are executed, allowing for a symbiotic contribution from both SGs and IBRs. Please note that the definition of control areas may change over time depending on the location of contingency and availability of resources. AGC balancing region, however are static. Therefore, IBRs would prefer to participate through APPF than directly in AGC which may need continuous re-tuning of the PI controllers in presence of reshuffling area boundaries. The benefit is that by combining APPF with AGC one can obtain a much better and faster transient performance of the generator frequencies than when using AGC only.

The voltage control problem is formulated using a similar argument with the difference that now we do not use multiple areas to execute an APPF. The hierarchies are defined by classifying the IBRs in the area in order of a sensitivity index for reactive power balance. SGs inside the contingent area are also counted in these hierarchies. APPF is run from one hierarchy to the next until the bus voltage at the contingent load bus is regulated to a steady-state value near the safe range of [0.95, 1.05] per unit (p.u.), while ensuring minimal disruption of voltages in the other buses inside this area. Numerical examples are shown for both control problems using a 33-bus, 3-area, 2-hierarchy power system model with multiple SGs and IBRs. The results, as indicated earlier, bring out two interesting points. First, they show that APPF prevents under-utilization of the renewables compared to when regular power flow is used for post-contingency setpoint dispatch. Second, they illustrate that combining APPF with AGC guarantees a much faster settling time for the frequency dynamics compared to when only AGC is used. Similar improvement in dynamic performance is seen for the voltage control problem as well.

The main contributions of the paper can be summarized as follows:
\begin{enumerate}
    \item We propose a hierarchical power flow solver referred to as APPF which decomposes the conventional power flow problem into smaller dimensional problems using physical hierarchies of a power system. APPF enables maximum utilization of IBRs, and limits the effect of disturbances to only those hierarchies that are in the vicinity of the disturbance source.
    \item The first variant of APPF aims to solve frequency regulation by generating appropriate power setpoints for the IBRs. This approach, unlike conventional AGC, does not need any frequency feedback to the generators. The regulation time also improves significantly due to the fast time constant of the IBR dynamics. 
    \item The second variant of APPF aims to solve voltage regulation by using hierarchies among different types of local VAR resources to cater to reactive power imbalance.
    \item Simultaneous application of APPF-based frequency and voltage control is also presented, where voltage control is prioritized over frequency control.
\end{enumerate}

Preliminary results on the frequency control problem were recently presented in our conference paper \cite{2020_Rahul_PESGM}. The results on frequency control reported here, however, are significantly extensive with a detailed theoretical formulation. The voltage control problem is completely new. The rest of the paper is organized as follows. Section \ref{freq_control_formulation} presents the problem set-up in terms of the hierarchical architecture of a given power system, followed by APPF-based frequency control. Section \ref{volt_control_formulation} extends the method to APPF-based hierarchical voltage control, and to simultaneous frequency and voltage control. Section \ref{simulation_result} presents numerical results. Section \ref{conclusion} concludes the paper.


\section{Problem I: Area-Prioritized Frequency Control}\label{freq_control_formulation}

\begin{figure}[h]
	\begin{center}
		\includegraphics[width=\linewidth]{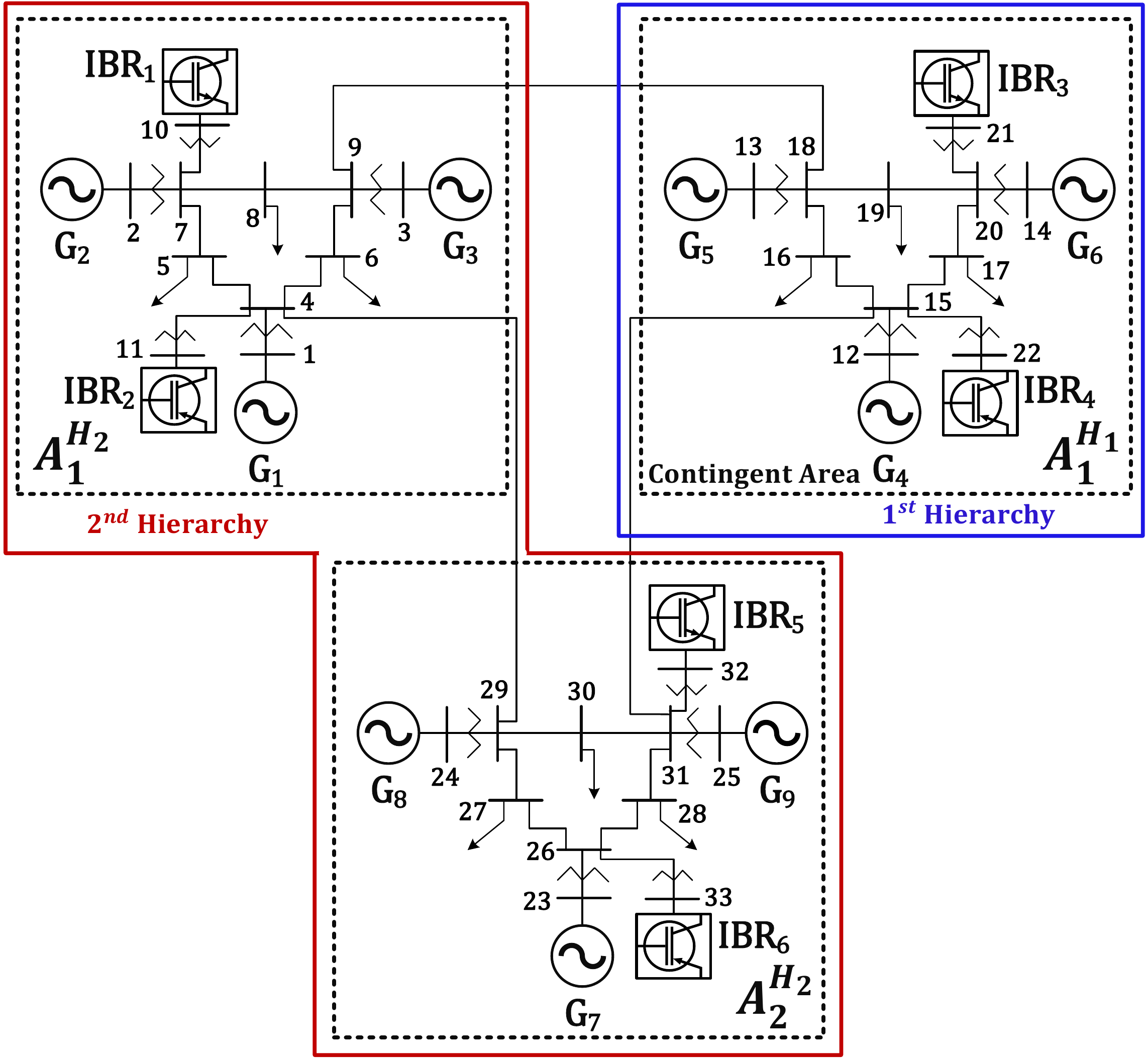}
		\caption{9-machine, 6-IBR, 33-bus power system}
		\label{33_Bus_Case_Study}
	\end{center}
\end{figure}

Consider a power system divided into a non-zero number of distinct control areas. As indicated earlier, the area where the contingency occurs is indexed as the first hierarchy, and subsequently the areas that share direct tie-line connections with it are collectively indexed as the second hierarchy, the ones that share connection with any area in the second hierarchy are collectively indexed as the third hierarchy, and so on. Accordingly, any area in the system is denoted as $A_j^{H_i}$ indicating that it is the $j^{th}$-area in the $i^{th}$-hierarchy. An example, which will be used later for simulations, is shown in Fig. \ref{33_Bus_Case_Study} in the form of a 33-bus power system model divided into 3 control areas forming 2-hierarchies. Each area consists of 3 SGs and 9 buses following \cite{2018_M_A_Pai_Chow} integrated with 2 IBR buses. Contingency occurs in area $A_1^{H_1}$ making it the $1^{st}$ hierarchy. The other two neighboring areas $A_1^{H_2}$, $A_2^{H_2}$ form the $2^{nd}$ hierarchy. The above concept of physical hierarchy can be extended to any power system divided into $n$ hierarchies. The $i^{th}$ hierarchy is divided into $n_i$ number of non-overlapping areas $\{A_{1}^{H_i}, A_{2}^{H_i},\ldots,A_{n_i}^{H_i}\}$. Note that, this decomposition is not dependent on any particular network topology. Also note that the term {\it hierarchy} here means  physical hierarchy, i.e., the physical location of the grid components, and not control hierarchy (for example, primary control, secondary control, and tertiary control) as commonly used in the power system control literature \cite{2012_Davoudi,2014_Nikos_H}.

Based on this power system architecture, our proposed frequency control strategy consists of three steps - (1) Detection of contingency, (2) headroom-based primary control, and finally (3) secondary control actuated sequentially from $1^{st}$ hierarchy to $n^{th}$ hierarchy following prioritization-based optimal dispatch. These three steps are described as follows.

\subsection{Detection of  active power imbalance} \label{detection_active_power}
The pre-contingency steady-state active power balance equation for the $i^{th}$ area in any hierarchy is written as
\begin{equation} \label{Power_balance_each_area}
\sum P_{load}^i(0^{-}) = \sum P_{tie}^i(0^{-}) + \sum P_{generation}^i(0^{-}),
\end{equation}
\noindent where, for any area $i$, $\sum P_{load}^i(0^{-})$ is the cumulative active power load, $\sum P_{generation}^i(0^{-})$ is the total active power generation, and $\sum P_{tie}^i(0^{-})$ is the total incoming active power from other areas. When a contingency happens, say in the form of a sudden unpredicted change of load, then the change in these three respective quantities will satisfy
\begin{equation} \label{del_Power_balance_each_area}
\sum \Delta P_{load}^i(t) = \sum \Delta P_{tie}^i(t) + \sum \Delta P_{generation}^i(t),
\end{equation}
for any time instant $t \geq 0$ where, $\sum \Delta P_{load}^i(t) = \sum P_{load}^i(t) - \sum P_{load}^i(0^{-})$. The same holds for tie-line and generation flows. We assume every area $i$ to be equipped with an area-level coordinator that receives sampled measurements of $P_{generation}^i(t)$ and $P_{tie}^i(t)$ streaming continuously over time from the generator buses and the tie-line buses. This can be achieved by placing high-resolution time-synchronized sensors such as Phasor Measurement Units (PMUs) to guarantee geometric observability of every generator and boundary buses of each individual area \cite{2018_Jain_Chakrabortty}. $P_{load}^i(t)$, however, may not necessarily be  measured. If the  contingency is due to tripping of any generator then the area coordinator can detect that directly from the measurement of  $\sum \Delta P_{generation}^i(t)$. If, however, the contingency is from an unforeseen load change then $\sum \Delta P_{load}^i \neq 0$ which implies that
\begin{equation} \label{Power_balance_contingent_area}
\Big|\sum \Delta P_{tie}^i(t)\Big| \neq \Big|\sum \Delta P_{generation}^i(t)\Big|,
\end{equation}
for every $t$. Thus, one simple way for area coordinators to detect the contingency would be to keep checking this condition at all times. If the coordinator of the $i^{th}$ area finds (3) to be true then it knows that the contingency has happened in its own area. For easy reference, we will refer to such an area as the {\it contingent} area, and the remaining areas as non-contingent areas. In reality, the generation capacity of a single area will be small compared to that of the grid, i.e., $\sum \Delta P_{generation}^i \approx$ 0. Therefore, the non-contingent areas will observe $\sum \Delta P_{tie}^i \approx$ 0, while the contingent areas will see $\sum \Delta P_{tie}^i \neq$ 0. The magnitude of the contingency can be measured as $\sum \Delta P_{load}^i \approx \sum \Delta P_{tie}^i$.

\subsection{Primary frequency control}\label{primary_freq_cont}
Once the contingency is detected and the contingent area is identified, the next step of our control strategy is to trigger primary control signals through the IBRs. The power dispatch for each IBR is chosen in proportion to its relative available capacity or headroom. The first step is to determine whether the contingent area in question is self-sufficient (i.e., whether the IBRs in the contingent area cumulatively have sufficient headroom to compensate for the load loss) or not. We define these checks as follows.

\noindent (1) $1^{st}$ hierarchy is \textbf{Self-sufficient} if 
\begin{equation} \label{self_sufficiency_condition_1st_hierarchy}
\sum \Delta P_{load}^{A_1^{H_1}} \leq \sum\limits_{j=1}^{N_{H_1}}h_j,
\end{equation}
\noindent where, $h_j$ is the available headroom in $j^{th}$ IBR of this hierarchy, and $N_{H_1}$ is the total number of IBRs in this hierarchy. The active power setpoint of the $j^{th}$ IBR is then updated according to its available headroom as
\begin{equation} \label{self_sufficiency_stpt_IBR_1st_hierarchy}
P_{IBR,j}^{\Upsilon} = P_{IBR,j}^{*} + ({h_j}/{\sum\limits_{j=1}^{N_{H_1}}h_j})\sum \Delta P_{load}^{A_1^{H_1}},
\end{equation}
\noindent $\forall j \in \{1,2,\ldots,N_{H_1}\}$. Where, $P_{IBR,j}^{*}$ is the pre-contingency setpoint and $P_{IBR,j}^{\Upsilon}$ is the solution of primary control for the $j^{th}$ IBR.

\noindent (2) $1^{st}$ hierarchy is \textbf{Self-deficient} if
\begin{equation} \label{self_defiiciency_condition_1st_hierarchy}
\sum \Delta P_{load}^{A_1^{H_1}} > \sum\limits_{j=1}^{N_{H_1}}h_j.
\end{equation}
\noindent In that case, the active power setpoint of the $j^{th}$ IBR in $A_1^{H_1}$ will be updated as
\begin{equation} \label{self_defiiciency_stpt_IBR_1st_hierarchy}
P_{IBR,j}^{\Upsilon} = P_{IBR,j}^{*} + h_j \qquad\forall j \in \{1,2,\ldots,N_{H_1}\}.
\end{equation}
Accordingly, the total deficit power $\Delta P_{req,_{H_1}}$ required by the $1^{st}$ hierarchy from the higher hierarchies is
\begin{equation} \label{self_defiiciency_deficit_1st_hierarchy}
\Delta P_{req,_{H_1}} = \sum \Delta P_{load}^{A_1^{H_1}} - \sum\limits_{j=1}^{N_{H_1}}h_j.
\end{equation}

Following the same logic as above, any $i^{th}$ hierarchy for $i>1$ is \textbf{Self-sufficient} if
\begin{equation} \label{self_sufficiency_condition_ith_hierarchy}
\Delta P_{req,_{H_{(i-1)}}} \leq \sum\limits_{k=1}^{N_{H_i}}h_k,\\
\qquad P_{A_j^{H_i}}^{\Upsilon} = ({\sum\limits_{k=1}^{N_{A_j^{H_i}}}h_k}/{\sum\limits_{k=1}^{N_{H_i}}h_k})\Delta P_{req,_{H_{(i-1)}}},
\end{equation}
\noindent$\forall j \in \{1,2,\ldots,n_i\}$, where, $\Delta P_{req,_{H_{(i-1)}}}$ is the deficit power required by $(i-1)^{th}$ hierarchy from $i^{th}$ hierarchy, $N_{A_j^{H_i}}$ is the number of IBRs in the $j^{th}$ area of the $i^{th}$ hierarchy, $P_{A_j^{H_i}}^{\Upsilon}$ is the corresponding power required from $j^{th}$ area of the $i^{th}$ hierarchy, and the total number of IBRs present in $i^{th}$ hierarchy = $N_{H_{i}} = \sum\limits_{j=1}^{n_i}N_{A_{j}^{H_{i}}}$. The active power setpoint of the $l^{th}$ IBR of the $j^{th}$ area in the $i^{th}$ hierarchy will be updated as
\begin{equation} \label{self_sufficiency_stpt_IBR_ith_hierarchy}
P_{IBR,l}^{\Upsilon} = P_{IBR,l}^{*} + ({h_l}/{\sum\limits_{k=1}^{N_{A_j^{H_i}}}h_k})P_{A_j^{H_i}}^{\Upsilon} \qquad\forall l \in \{1,2,\ldots,N_{A_j^{H_i}}\}.
\end{equation}
Similarly, the $i^{th}$ hierarchy ($i\neq1$) is \textbf{Self-deficient} if 
\vspace{-2pt}
\begin{equation}
\Delta P_{req,_{H_{(i-1)}}} > \sum\limits_{k=1}^{N_{H_i}}h_k,
\end{equation}
with the active power setpoint of $l^{th}$ IBR of $j^{th}$ area in $i^{th}$ hierarchy updated as
\vspace{-2pt}
\begin{equation} \label{self_defiiciency_stpt_IBR_ith_hierarchy}
P_{IBR,l}^{\Upsilon} = P_{IBR,l}^{*} + h_l \qquad\forall l \in \{1,2,\ldots,N_{A_j^{H_i}}\}.
\end{equation}
Therefore, the amount of deficit power ($\Delta P_{req,_{H_i}}$) required by $i^{th}$ hierarchy from $(i+1)^{th}$ hierarchy is as follows:

\vspace{-12pt}
\begin{equation} \label{self_defiiciency_deficit_ith_hierarchy}
\Delta P_{req,_{H_i}} = \Delta P_{req,_{H_{(i-1)}}} - \sum\limits_{k=1}^{N_{H_i}}h_k.
\end{equation}
The coordinator of the control areas under the $1^{st}$ hierarchy solves (\ref{self_sufficiency_stpt_IBR_1st_hierarchy}) and (\ref{self_defiiciency_stpt_IBR_1st_hierarchy}) while those under the $i^{th}$ hierarchy solve (\ref{self_sufficiency_stpt_IBR_ith_hierarchy}) and (\ref{self_defiiciency_stpt_IBR_ith_hierarchy}) to generate the respective active power setpoints, which are then communicated to the IBRs in their respective areas. The setpoints, thereafter, are actuated through the  Voltage Source Converter (VSC) based model of the IBRs \cite{2014_NREL_IBR_Model}.

\subsection{Secondary frequency control}\label{secondary_freq_cont}
Primary control alone, however, cannot revive the system-wide frequency to the synchronous value (60 Hz in the United States). We, therefore, next propose a new method for executing secondary control that is based on the notion of area-prioritization. We introduce a new setpoint calculation method referred to as Area-Prioritized Power Flow (APPF) that generates power setpoints for the IBRs in each area, but in a sequential way of preference from one hierarchy to the next, starting from the contingent area. The goal is to minimize the effect of the contingency from spreading too far beyond its source through prioritized resource utilization. This is done by solving a local power flow for the contingent area $A_1^{H_1}$ as the first step, and then using the solution to solve the local power flow of the next hierarchy, and so on, resulting in a $n$-stage hierarchical optimization problem. 

Considering the pre-contingency power flow solution ($x^{*}$) as the initial condition, stage $i$ of APPF targets the buses of the $i^{th}$ hierarchy, and minimizes the deviation in the tie-line power flowing between $i^{th}$ and $(i+1)^{th}$ hierarchy to support the power required by the $(i-1)^{th}$ hierarchy. Additionally, it also tries to minimize the deviation of the IBR active power setpoints ($P_{IBR,k}$) from their dispatched primary control setpoints ($P_{IBR,k}^{\Upsilon}$) so that the IBR does not experience a significant jump in its actuation. This, in turn, reduces the transient peaks in the frequency trajectories when secondary control overwrites the primary control setpoints. The mathematical formulation of APPF is presented as follows:

\noindent $\bullet$ \textbf{Objective function of $i^{th}$ stage:}
\begin{equation} \label{Objective_stage_i}
\begin{aligned}
x_i^{**} = \underset{x}{\textrm{argmin}} \quad w_1 .\left\|\begin{bmatrix}
\sum\limits_{k=1}^{r_{i(i+1)}} P_{tie,k}^{i(i+1)} - \sum\limits_{k=1}^{r_{i(i+1)}} P_{tie,k}^{{i(i+1)}^{*}}\\
\vspace{-5pt}\\
\sum\limits_{k=1}^{r_{i(i+1)}} Q_{tie,k}^{i(i+1)} - \sum\limits_{k=1}^{r_{i(i+1)}} Q_{tie,k}^{{i(i+1)}^{*}}\end{bmatrix}\right\|^2\\
+ w_2 .\sum\limits_{k=1}^{N_{H_i}} \left\|P_{IBR,k} - P_{IBR,k}^{\Upsilon}\right\|^2,
\end{aligned}
\end{equation}

\noindent where, $x$ denotes the free optimization variables and $x_i^{**}$ is the solution of $i^{th}$ stage of APPF. The definition of $x$ for different bus types are listed in Table \ref{chart_of_variables}. $r_{i(i+1)}$ is the number of tie lines between $i^{th}$ and $(i+1)^{th}$ hierarchy. $P_{tie,k}^{i(i+1)}$ and $Q_{tie,k}^{i(i+1)}$ are the active and reactive power flows through the $k^{th}$ tie line whose pre-contingency values are $P_{tie,k}^{{i(i+1)}^{*}}$ and $Q_{tie,k}^{{i(i+1)}^{*}}$ respectively. $w_1$ and $w_2$ represent the corresponding weights of the objectives. Please note that the two objectives should be weighed according to the system operator's preference for prioritizing the speed of recovery over causing additional transients due to rapid adjustments of the IBR setpoints.
\begin{table}[h]  
    \centering 
    \caption{Chart of Variables for $i^{th}$ Stage Optimization}
    \label{chart_of_variables}
    \vspace{-5pt} 
    \begin{tabular}{ | c | c | c | c | c |}
    \hline
    \backslashbox{\textbf{Bus associated with}}{\textbf{Bus variables}}& $|V|$ & $\theta$ & P & Q\\ \hline
    Sync-gen & ${\color{red}\Huge{*}}$ & ${\color{blue}\Huge{\uplus}}$ & ${\color{red}\Huge{*}}$ & ${\color{blue}\Huge{\uplus}}$\\ \hline
	Load & ${\color{blue}\Huge{\uplus}}$ & ${\color{blue}\Huge{\uplus}}$ & ${\color{red}\Huge{*}}$ & ${\color{red}\Huge{*}}$\\ \hline
    No component & ${\color{blue}\Huge{\uplus}}$ & ${\color{blue}\Huge{\uplus}}$ & ${\color{red}\Huge{*}}$ & ${\color{red}\Huge{*}}$\\ \hline
    IBR & ${\color{red}\Huge{*}}$ & ${\color{red}\Huge{*}}$ & ${\color{blue}\Huge{\uplus}}$ & ${\color{blue}\Huge{\uplus}}$\\ \hline
    Tie-line to $(i+1)^{th}$ hierarchy & ${\color{red}\Huge{*}}$ & ${\color{red}\Huge{*}}$ & ${\color{blue}\Huge{\uplus}}$ & ${\color{blue}\Huge{\uplus}}$\\ \hline
    Tie-line to $(i-1)^{th}$ hierarchy ($i\neq1$) & ${\color{blue}\Huge{\uplus}}$ & ${\color{blue}\Huge{\uplus}}$ & ${\color{red}\Huge{*}}$ & ${\color{red}\Huge{*}}$\\ \hline
    \end{tabular}\\
    \vspace{2mm} 
  ${\color{red}\Huge{*}}$ = fixed, ${\color{blue}\Huge{\uplus}}$ = free optimization variables $x$,\\
  $|V|$ and $\theta$ are the phasor voltage magnitude and angle respectively. $P$ and $Q$ are the active and reactive power injections to the bus respectively.
\end{table}\\
\vspace{-20pt}

\begin{figure}[h]
	\begin{center}
		\includegraphics[width=9cm]{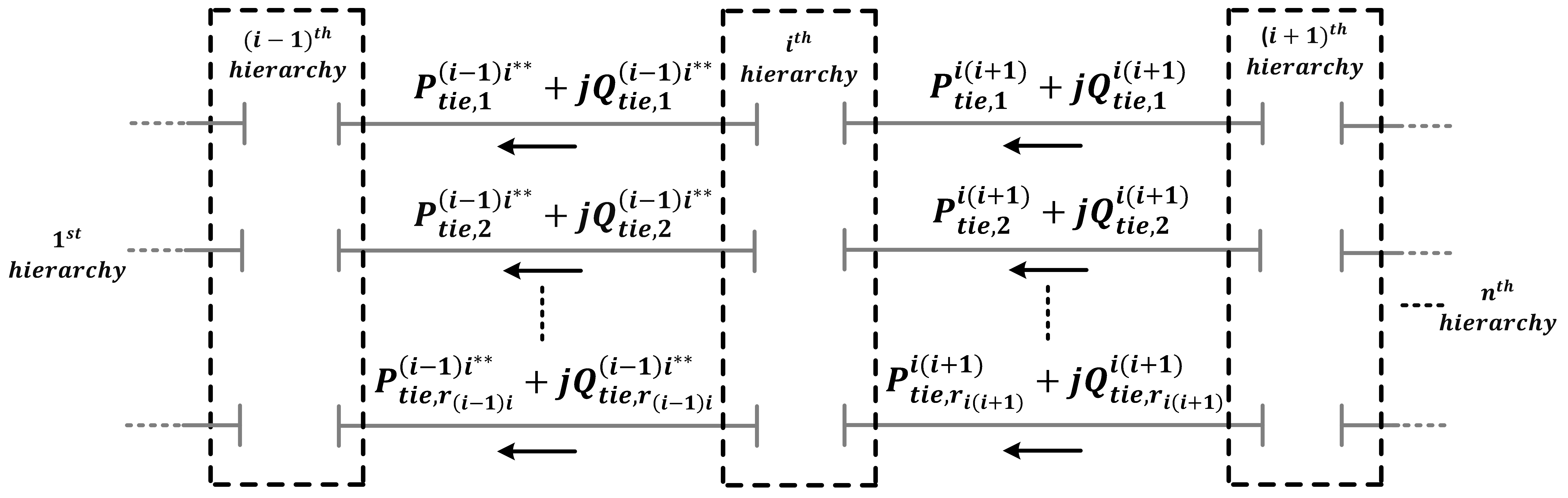}
		\vspace{-10pt}
		\caption{Power flow information needed for optimization in the $i^{th}$ hierarchy}
		\label{ith_Stage_Optimization}
	\end{center}
\end{figure}

\noindent $\bullet$ \textbf{Equality constraints:} The power balance in the $i^{th}$ hierarchy can be represented by
\begin{equation} \label{Power_Balance_stage_1}
\qquad 0 = (\mathbf{Y}^{i}\mathbf{V}^{i})^{*}\circ\mathbf{V}^{i} - (\mathbf{P}^{i}+j\mathbf{Q}^{i}) ,
\end{equation}
\noindent where, $\mathbf{Y}^{i} \in \mathbb{C}^{k^i \times k^i}$ is the admittance matrix of the $i^{th}$ hierarchy, $\circ$ is the element-wise multiplication, $*$ is the element-wise complex conjugate operator, and $k^i$ is the number of buses in the $i^{th}$ hierarchy. The bold symbols $\mathbf{V}^{i}$, $\mathbf{P}^{i}$ and  $\mathbf{Q}^{i}$ denote vector quantities whose expressions are written as follows. Let  $S^i$ be the set of buses in $i^{th}$ hierarchy, and $B^{i(i+1)}$ denote the set of boundary buses in the areas of $i^{th}$ hierarchy connecting $i^{th}$ hierarchy with $(i+1)^{th}$ hierarchy. For the $1^{st}$ hierarchy (i.e., $i=1$) :\\
\vspace{-5pt}
\begin{equation} \label{Eq_fre1}
\noindent\mathbf{V}^{i}=\begin{bmatrix}
\begin{bmatrix}\mathbf{V}_{\alpha_1}\end{bmatrix}_{r_{12}\times1}\\
\begin{bmatrix}\mathbf{V}_{\alpha_2}\end{bmatrix}_{(k^1-r_{12})\times1}\\
\end{bmatrix}, \qquad \alpha_1 \in B^{12},\\
\end{equation}
\begin{equation}\label{Eq_fre2}
\mathbf{P}^{i}+j\mathbf{Q}^{i}=\begin{bmatrix}
\begin{bmatrix}
P_{tie,1}^{12} + j Q_{tie,1}^{12}\\
P_{tie,2}^{12} + j Q_{tie,2}^{12}\\
\vdots\\
P_{tie,r_{12}}^{12} + j Q_{tie,r_{12}}^{12}\\
\end{bmatrix}\\
\begin{bmatrix}\mathbf{P}_{\alpha_2}+j\mathbf{Q}_{\alpha_2}\end{bmatrix}_{(K^1-r_{12})\times1}\\
\end{bmatrix}, \qquad \alpha_2 \in \{S^1 - B^{12}\},\\
\end{equation}
where, each element of $\mathbf{V}_{\alpha_1}$ and $\mathbf{V}_{\alpha_2}$ means $|V|\phase{\theta}$. Similarly, for any $i^{th}$ hierarchy where $i\neq1$, the areas of the $i^{th}$ hierarchy tries to supply the required power to the $(i-1)^{th}$ hierarchy, resulting in



\begin{equation}\label{Eq_fre3}
\mathbf{V}^{i}=\begin{bmatrix}
\begin{bmatrix}\mathbf{V}_{\alpha_1}\end{bmatrix}_{r_{(i-1)i}\times1}\\
\begin{bmatrix}\mathbf{V}_{\alpha_2}\end{bmatrix}_{r_{i(i+1)}\times1}\\
\begin{bmatrix}\mathbf{V}_{\alpha_3}\end{bmatrix}_{(k^i-r_{(i-1)i}-r_{i(i+1)})\times1}
\end{bmatrix},
\end{equation}
where, $\alpha_1 \in B^{(i-1)i}, \alpha_2 \in B^{i(i+1)}, \quad \alpha_3 \in \{S^i - B^{(i-1)i} - B^{i(i+1)}\}$.
\begin{equation}\label{Eq_fre4}
\mathbf{P}^{i}+j\mathbf{Q}^{i}=\begin{bmatrix}
\begin{bmatrix}\mathbf{P}+j\mathbf{Q}\end{bmatrix}^{i-1}_{r_{(i-1)i}\times1}\\
\begin{bmatrix}\mathbf{P}+j\mathbf{Q}\end{bmatrix}^{i+1}_{r_{i(i+1)}\times1}\\
\begin{bmatrix}\mathbf{P}_{\alpha_3}+j\mathbf{Q}_{\alpha_3}\end{bmatrix}_{(k^i-r_{(i-1)i}-r_{i(i+1)})\times1}
\end{bmatrix},
\end{equation}
\vspace{-5pt}
\noindent\begin{multline}\label{Eq_fre4_1}
\begin{bmatrix}\mathbf{P}+j\mathbf{Q}\end{bmatrix}^{i+1}\\=\begin{bmatrix}
(P_{tie,1}^{i(i+1)} + j Q_{tie,1}^{i(i+1)}) \cdots (P_{tie,r_{i(i+1)}}^{i(i+1)} + j Q_{tie,r_{i(i+1)}}^{i(i+1)})\\
\end{bmatrix}^T,
\end{multline}
and
\noindent\begin{multline}\label{Eq_fre5}
\begin{bmatrix}\mathbf{P}+j\mathbf{Q}\end{bmatrix}^{i-1}\\=-\begin{bmatrix}
(P_{tie,1}^{{(i-1)i}^{**}} + j Q_{tie,1}^{{(i-1)i}^{**}}) \cdots (P_{tie,r_{(i-1)i}}^{{(i-1)i}^{**}} + j Q_{tie,r_{(i-1)i}}^{{(i-1)i}^{**}})
\end{bmatrix}^T
\end{multline}
is the vector from the solution of $(i-1)^{th}$ optimization stage. Fig. \ref{ith_Stage_Optimization} shows the information flow between hierarchies for executing $i^{th}$ optimization stage. The tie-line losses are assumed to be negligible in the above formulation. To summarize, equations (\ref{Power_Balance_stage_1}) - (\ref{Eq_fre5}) form the equality constraints for the $i^{th}$ stage of APPF.\\
\vspace{-10 pt}

\noindent $\bullet$ \textbf{Inequality constraints:}
We impose operational constraints on the bus voltage magnitudes, active and reactive power outputs of generation resources as follows: 
\begin{equation} \label{V_Bus_limit}
|V|_{j}^{min} \leq |V|_{j} \leq |V|_{j}^{max},
\end{equation}
where, $|V|_{j}$ is the $j^{th}$ bus voltage magnitude. We consider $|V|_{j}$ to be bounded between 0.95 p.u. and 1.05 p.u. \cite{1994_Kundur_Book},
\begin{equation} \label{P_SG_limit}
\begin{cases}
P_{SG,j}^{min} \leq P_{SG,j} \leq P_{SG,j}^{max}, \\
Q_{SG,j}^{min} \leq Q_{SG,j} \leq Q_{SG,j}^{max},
\end{cases}   
\end{equation}
where, $P_{SG,j}$ and $Q_{SG,j}$ are active and reactive power outputs of $j^{th}$ SG respectively,
\begin{equation} \label{P_IBR_limit}
\begin{cases}
P_{IBR,j}^{min} \leq P_{IBR,j} \leq P_{IBR,j}^{max}, \\
Q_{IBR,j}^{min} \leq Q_{IBR,j} \leq Q_{IBR,j}^{max},
\end{cases}   
\end{equation}
where, $P_{IBR,j}$ and $Q_{IBR,j}$ are active and reactive power outputs of $j^{th}$ IBR respectively. The lower bound $(\cdot)^{min}$ and the upper bound $(\cdot)^{max}$ of variables in equations (\ref{V_Bus_limit}) - (\ref{P_IBR_limit}) determine the feasible solution space for the optimization. Active power flows through the $k^{th}$ tie-line connecting $i^{th}$ hierarchy with $(i+1)^{th}$ hierarchy are also bounded by their corresponding thermal rating and the magnitude of the contingency as
\begin{equation} \label{P_tie_limit}
P_{tie,k}^{i(i+1)} \leq P_{tie,k}^{i(i+1)^*} + \Delta P_{load}^{A_1^{H_1}} \leq P_{tie,k}^{i(i+1)}\Big|_{thermal}.
\end{equation}
\noindent Equations (\ref{V_Bus_limit}) - (\ref{P_tie_limit}) form the inequality constraints for the $i^{th}$ stage of APPF. The coordinator of the control areas under the $i^{th}$ hierarchy solve (\ref{Objective_stage_i})-(\ref{P_tie_limit}) to generate the respective active power setpoints, which are then communicated to the IBRs in their respective areas. The setpoints, as usual, are actuated through the VSC based model of the IBRs \cite{2014_NREL_IBR_Model}. As the time constants of the VSC dynamics are smaller compared to that of AGC \cite{2021_Duncan}, the speed of frequency regulation improves significantly. This will be seen in our simulation results in section \ref{simulation_result}.


\section{Problem II: Area-Prioritized Voltage Control}\label{volt_control_formulation}
Unlike problem I, where the aim is to revive the frequency of the buses to the synchronous frequency, the goal of voltage control is to maintain the voltage magnitude at the buses within a desired band by proper dispatch of the reactive power setpoints of local area-level resources (both IBRs and SGs). Accordingly, we modify the definition of the hierarchies for this problem as follows. We prepare a ranked list of IBRs, and thereafter the SGs in the system depending on how much impact they have on the voltage regulation for the contingent bus. This ranking dictates the prioritization order of the IBRs as well as for the SGs, thereby defining the hierarchies from one resource to the next. The control algorithm involves four steps, as explained next.

\subsection{Detection of reactive power imbalance}\label{detection_reactive_power}
In pre-contingency steady-state, the consecutive samples in the streaming measurements of any bus voltage at any time point $t$ satisfies
\begin{equation} \label{pre_cont_V_monitoring}
|V|_{pre-contingency}-\beta_1 \leq |V|_{t} \leq |V|_{pre-contingency}+\beta_2,
\end{equation}
where, $\beta_1$, $\beta_2$ are design specific voltage deviation constants which are, usually taken as $5\%$ of the base voltage. After the contingency, this relation may change to $|V|_{t} < |V|_{pre-contingency}-\beta_1$ or $|V|_{t} > |V|_{pre-contingency}+\beta_2$, which can enable the area coordinator to detect the contingency. Following this detection, the amount of reactive power imbalance is calculated by the area-level state estimator.

\subsection{Sensitivity coefficient-based IBR ranking}
The next step is to rank the IBRs in the control area with respect to the contingent bus. As shown in \cite{2016_Mani_Voltage}, the bus voltage variation is related to the reactive power injection through sensitivity coefficients $dV/dQ$ as follows:
\begin{equation} \label{dV_dQ_Sensitivity_eq_1}
\begin{bmatrix}
\Delta V_1 \\ \Delta V_2 \\ \vdots \\ \Delta V_{n_j}
\end{bmatrix} = \begin{bmatrix}
\frac{dV_1}{dQ_1} & \frac{dV_1}{dQ_2} & \cdots & \frac{dV_1}{dQ_{n_j}}\\
\frac{dV_2}{dQ_1} & \frac{dV_2}{dQ_2} & \cdots & \frac{dV_2}{dQ_{n_j}}\\
\vdots & \vdots & \vdots & \vdots\\
\frac{dV_{n_j}}{dQ_1} & \frac{dV_{n_j}}{dQ_2} & \cdots & \frac{dV_{n_j}}{dQ_{n_j}}\\
\end{bmatrix}\begin{bmatrix}
\Delta Q_1 \\ \Delta Q_2 \\ \vdots \\ \Delta Q_{n_j}
\end{bmatrix},
\end{equation}
\begin{equation} \label{dV_dQ_Sensitivity_eq_2}
\Longrightarrow[\Delta \mathbf{V}]=[\mathbf{S}][\Delta \mathbf{Q}],
\end{equation}
where, $n_j$ is the total number of buses in the $j^{th}$ area. We use the sensitivity matrix $[\mathbf{S}]$ to determine which IBRs can affect the contingent bus voltage most effectively in post-contingency condition. The sensitivity coefficients are used to rank the IBRs and group them using following steps. Say, contingency happened at $k^{th}$ load bus of the $j^{th}$ control area. The contingent bus voltage deviation relates to the reactive power injections as follows:
\vspace{-5pt}
\begin{equation} \label{dV_dQ_Sensitivity_eq_2}
\Delta \mathbf{V}(k,1)=[\mathbf{S}(k,1:n_j)][\Delta \mathbf{Q}(1:n_j,1)],
\end{equation}
From $[\mathbf{S}(k,1:n_j)]$ array, the coefficients corresponding to IBR buses are identified and the IBRs are ranked from highest sensitivity to the lowest. The implicit assumption here is that the area goes through a smaller magnitude of reactive power imbalance compared to its total capacity while operating in the safe region of the nose curve, i.e., the contingency does not push the system to its voltage instability limits \cite{1990_V_Q_sense,1998_Voltage_Stability_Book}. This will ensure the validity of sensitivity coefficient based study around  the  actual  operating point of the power system.

\subsection{Primary voltage control}\label{primary_volt_cont}
Once the IBRs are ranked, the cumulative reactive power headrooms (i.e., maximum reactive power injection capability) of highly ranked IBRs are compared with the amount of reactive power imbalance to determine the minimal group of IBRs that need to participate in primary control. We refer to this set as $IBR_{class1}$ and the rest as $IBR_{class2}$. If $h_{k}$ is the available  reactive power headroom of the $k^{th}$-ranked IBR, then to satisfy an imbalance of $\Delta Q_L$, the IBRs in class1 must satisfy
\vspace{-7 pt}
\begin{equation}\label{reactive_headroom_compare}
\Delta Q_L \leq \sum\limits_{k=1}^{N_1}h_k,\\
\end{equation}
where, $N_{1}$ is the number of IBRs in class1. The reactive power setpoints in class1 are then updated by sequentially utilizing the headrooms of the IBRs following the ranking order as
\vspace{-5 pt}
\begin{equation} \label{primary_reactive_dispatch_eq1}
Q_{IBR,j}^{\Upsilon} = Q_{IBR,j}^{*} + h_j \qquad\forall j \in \{1,2,\ldots,(N_1-1)\}.
\end{equation}
\vspace{-10 pt}
\begin{equation} \label{primary_reactive_dispatch_eq1}
Q_{IBR,N_{1}}^{\Upsilon} = Q_{IBR,N_1}^{*} + (\Delta Q_L - \sum\limits_{k=1}^{N_1-1}h_k)
\end{equation}

\subsection{Secondary voltage control}
The headroom based setpoint dispatch, however, may result in high reactive power injection at the IBR buses that, in turn, may push their voltages outside the allowable range of $[0.95,\, 1.05]$ p.u. together with their neighboring buses. Therefore, the optimization objective of the secondary control is to involve all IBRs and conventional dynamic reactive power reserves such as SGs categorically, and achieve an overall improvement of the voltage profile of the contingent area. A common operational requirement is that the load bus voltage must lie within a tight bound, typically $5\%$ around the reference voltage. Accordingly, the objective of secondary level optimization is as follows:

\noindent $\bullet$ \textbf{Objective function:}
\vspace{-2pt}
\begin{equation} \label{Voltage_Objective}
\begin{aligned}
x^{**} = \underset{x}{\textrm{argmin}} \sum\limits_{k\in n_{L_j}}|||V_k| - 1||^2,
\end{aligned}
\end{equation}
where, $x$ denotes the free variables for the optimization steps as shown in algorithm 1. $x^{**}$ is the solution of sequential APPF and $n_{L_j}$ is the number of load buses in the contingent area. The network power balance equation (\ref{Power_Balance_stage_1}) forms the equality constraint. The inequality constraints are same as equations (\ref{V_Bus_limit}) - (\ref{P_IBR_limit}). However, in this case, we allow voltage relaxation to the highly ranked IBRs and sequentially involve conventional reactive power reserves of SGs to attain the desired voltage profile. The detailed steps of sequential optimization to gradually involve and utilize different categories of reactive power resources are described in algorithm 1.

Note that the SGs start participating in reactive power compensation step 5 onward, after the available capacities of $IBR_{class1}$ are fully exhausted in previous steps. The automatic voltage regulator (AVR) setpoints are adjusted to dispatch the required reactive power from SGs. The adaptive initialization used in algorithm 1 directs the solution to a local minima. If the $IBR_{class1}$, SGs, and $IBR_{class2}$ have enough cumulative capacity to compensate for the reactive power imbalance, a feasible solution will be obtained at the end of this sequential optimization while maintaining the network constraints.

\subsection{Simultaneous Frequency and Voltage Control}\label{freq_volt_control_together_algo}

The hierarchical frequency and voltage control designs proposed so far implicitly assume that the frequency and voltage dynamics have weak coupling so that each control algorithm can be designed independently of the other \cite{2013_Johansson_Distributed_vs_Centralized, 2013_Sezi_synchronization}.  Although simultaneous imbalance of active and reactive

\noindent\fbox{\parbox{\linewidth}{
\noindent \textbf{Algorithm 1: Sequential APPF for Voltage Control}\\
\vspace{-3pt}
\noindent\rule{8.8cm}{0.4pt}\\
\vspace{-0.5pt}
\noindent\textbf{Step 1:} Initialization with pre-contingency equilibrium.\\
\vspace{1pt}
\noindent\textbf{Step 2:} Set local bound of bus voltages as $[0.95,\, 1.05]$ p.u. and global bounds around pre-contingency equilibrium as:\\
\vspace{-0.5pt}
\qquad\quad($|V|^{max}, |V|^{min}$) = $|V|_{pre-contingency}\pm\Delta \upsilon$,\\
where, $\Delta \upsilon$ is a design parameter based on the maximum safe bound of network operation.\\
\vspace{1pt}
\noindent\textbf{Step 3:} Solve secondary level optimization with objective (\ref{Voltage_Objective}) with the following conventions on the bus variables:\\
\vspace{1pt}
\mybox{gray}{SG bus: $P$,$Q \rightarrow $ fixed, $|V|$,$\theta \rightarrow $ free,\\
\vspace{-0.5pt}
$IBR_{class1}$, $IBR_{class2}$ bus: $|V|$,$\theta \rightarrow $ fixed, $P$, $Q \rightarrow $ free,\\
\vspace{-0.5pt}
Load bus, transfer bus: $P$, $Q \rightarrow $ fixed, $|V|$, $\theta \rightarrow $ free,}\\
\vspace{1pt}
The above fixed variables are held at initialized value.\\
\vspace{1pt}
\noindent\textbf{Step 4:} If solution of Step 3 is non-convergent then relax the voltages of SG, IBR, and transfer buses to their global bound ($|V|^{max}, |V|^{min}$). Repeat Step 3 and then go to Step 5.\\
\vspace{1pt}
\noindent\textbf{Step 5:} Use the solution from Step 3 as initialization to repeat the secondary level optimization. The conventions on the bus variables are as follows:\\
\vspace{1pt}
\mybox{gray}{SG bus: $\theta$, $P \rightarrow $ fixed, $|V|$, $Q \rightarrow $ free,\\
\vspace{-0.5pt}
$IBR_{class1}$ bus: $P$,\,$Q \rightarrow $ fixed, $|V|$, $\theta \rightarrow $ free,\\
\vspace{-0.5pt}
$IBR_{class2}$ bus: $|V|$, $\theta \rightarrow $ fixed, $P$, $Q \rightarrow $ free,\\
\vspace{-0.5pt}
Load bus, transfer bus: $P$, $Q \rightarrow $ fixed, $|V|$, $\theta \rightarrow $ free,}\\
\vspace{1pt}
\noindent\textbf{Step 6:} If solution of step 5 is still non-convergent then use the solution from Step 5 as initialization to repeat the secondary level optimization following the convention:\\
\vspace{1pt}
\mybox{gray}{SG bus: $P$, $Q \rightarrow $ fixed, $|V|$,$\theta \rightarrow $ free,\\
\vspace{-0.5pt}
$IBR_{class1}$ bus: $P$, $Q \rightarrow $ fixed, $|V|$, $\theta \rightarrow $ free,\\
\vspace{-0.5pt}
$IBR_{class2}$ bus: $|V|$, $\theta \rightarrow $ fixed, $P$, $Q \rightarrow $ free,\\
\vspace{-0.5pt}
Load bus, transfer bus: $P$, $Q \rightarrow $ fixed, $|V|$, $\theta \rightarrow $ free.}}}\\

\noindent power  may happen rarely, it is not an impossibility. Therefore, we next list a possible solution for simultaneous frequency and voltage control by combining the formulations in sections \ref{freq_control_formulation} and \ref{volt_control_formulation}. As both APPFs try to utilize the available MVA capacity of IBRs in terms of active or reactive power headrooms, one control algorithm has to be prioritized over the other to resolve any arising conflict.

Voltage control can be provided by local VAR resources. Therefore, when both active and reactive injections are needed, area-coordinators can prioritize reactive power injections first from the available headroom of the IBRs near the contingent bus. If an IBR still has any headroom left for active power in addition to its reactive power headroom, it will participate in frequency control. The local area can lean on the neighboring areas for any deficit real power as described in section \ref{freq_control_formulation}. The prioritization of voltage control over frequency control is done to prevent local voltage violations first before moving to the system-wide control of frequency.

\begin{figure}[h]
	\begin{center}
		\includegraphics[width=9cm]{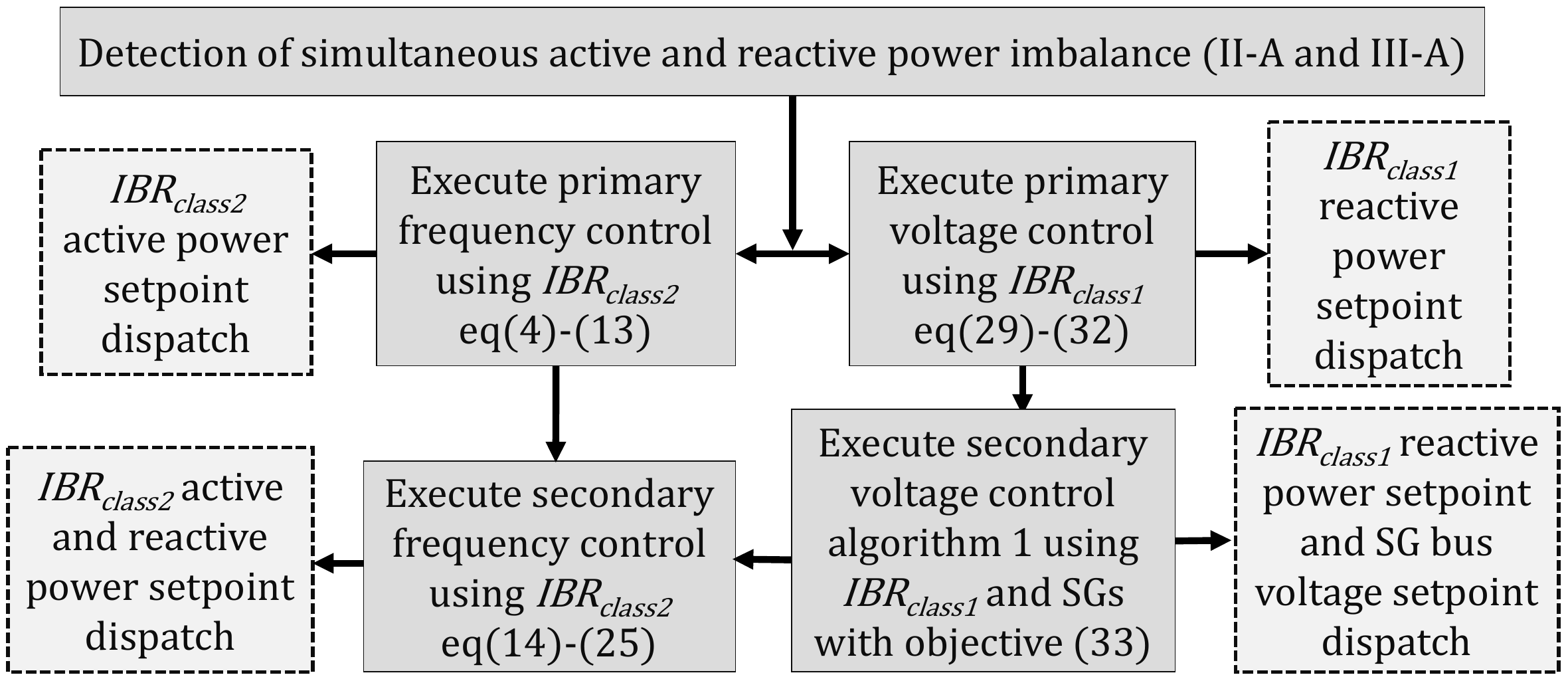}
		\vspace{-10pt}
		\caption{Flow-chart showing the hierarchical execution of actions for simultaneous frequency and voltage control}
		\label{Freq_Volt_Flow_Chart}
	\end{center}
\end{figure}

Following the simultaneous triggering of detection phases \ref{detection_active_power} and \ref{detection_reactive_power}, primary voltage control can be executed using reactive power headrooms of $IBR_{class1}$. Thereafter, the conventional VAR sources such as SGs can compensate for the residual reactive power requirements following Algorithm 1. Consequently, the IBRs of $IBR_{class2}$ will be available for utilization in primary and secondary layers of hierarchical frequency control design following the formulations in \ref{primary_freq_cont} and \ref{secondary_freq_cont}. Hierarchical secondary frequency control will also require reactive power setpoint dispatch for $IBR_{class2}$ to compensate for the new network losses. The flow-chart in Fig. \ref{Freq_Volt_Flow_Chart} demonstrates the steps for APPF-based simultaneous control of frequency and voltage with priority  given to voltage control.\\
\vspace{-20pt}

\vspace{-15pt}

\noindent\textbf{Remark 1:} The sensitivity coefficients in (\ref{dV_dQ_Sensitivity_eq_2}), and the IBR headrooms in (\ref{self_sufficiency_condition_1st_hierarchy})-(\ref{self_defiiciency_deficit_ith_hierarchy}) and (\ref{reactive_headroom_compare})-(\ref{primary_reactive_dispatch_eq1}) that serve as prior information for APPF do not need to be computed in real-time after the contingency. The area coordinators will have sufficient time to keep these information ready before any contingency hits.

\noindent\textbf{Remark 2:} The execution time for APPF is dictated solely by the size of the contingent area, irrespective of how large the power system is. Partitioning the grid into a large number of small-sized areas will result in quicker power flow solutions per hierarchy, but the execution has to wait till the calculations from all the participating hierarchies are complete. On the other hand, dividing the grid into smaller number of large-sized areas will involve lesser amount of waiting time from one hierarchy to another, but larger time for solving power flow per hierarchy. System designer must design the hierarchies considering these points for their specific power system of interest.

\noindent\textbf{Remark 3:} Note that APPF will only be applied in response to a contingency. For regular rolling changes in loads and generations, the operator may resort back to conventional or Regular Power Flow (RPF) to update the IBR setpoints. This will also ensure that no IBR gets saturated at its maximum or minimum operating limits when APPF is applied for the next contingency. 

\noindent\textbf{Remark 4:} APPF can address multiple contingencies as a single event if they are detected simultaneously in the same control area or in the areas of consecutive hierarchies. However, if another contingency in the same area or in a distant area overlaps with the execution stages of APPF from last contingency, the traditional AGC has to address the second event as the APPF responding to the first contingency can not be suspended midway.

\vspace{-5pt}
\section{Cyber-Physical Architecture}\label{CPS_architecture}
\begin{figure*}[!b]
	\centering
	\includegraphics[width=\textwidth]{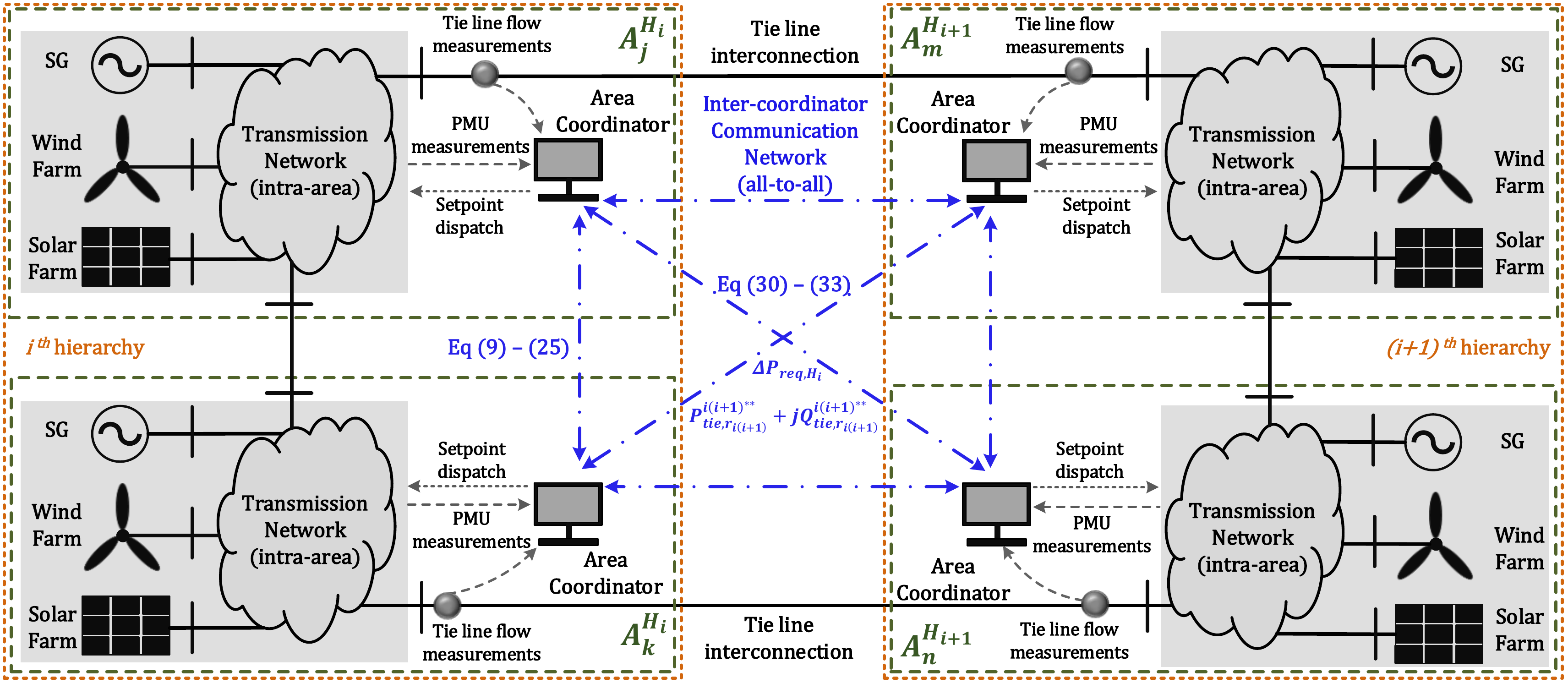}
	\caption{Cyber-physical architecture for implementing APPF}
	\label{CPS_Data_flow_Diagram}
\end{figure*}

Fig. \ref{CPS_Data_flow_Diagram} shows the cyber-physical (CPS) infrastructure to execute the proposed hierarchical frequency and voltage control. Each area ($A_j^{H_i}$) has its own monitoring devices and an area-coordinator, as mentioned earlier. The area coordinator consists of a local state estimator and a setpoint computer. It is assumed that sufficient number of sensors are available in the area to provide measurements covering geometric observability of the states so that the state estimator can represent the most updated real-time model of the area. For example, an exhaustive list of sensors can be listed as PMUs at boundary buses to measure tie-line power flows, PMUs at SG and IBR buses to measure bus voltages and active-reactive power outputs. Ideally to achieve full observability, it is preferred for every substation in the area should have at least one PMU.

In terms of communication, a Wide Area Network (WAN) is needed to connect the boundary buses,  the SG buses, and the IBR buses to their area coordinators. Local Area Network (LAN) or SCADA communication is needed between the monitoring devices and area coordinator. A Wide Area Network (WAN) is also needed between the area coordinators for periodic and event-triggered communication so that they can update the cumulative headrooms of each area as well as exchange setpoint information between the consecutive hieararchies. For frequency control, the communication links from the tie-line flow sensors and the inter-area communications are active, whereas, those from the SGs are inactive. For voltage control, it is the opposite. All other communication links depicted in Fig. \ref{CPS_Data_flow_Diagram} are utilized in both frequency and voltage control. Note that for frequency control, the physical hierarchies share only the tie-line power demands with each other. Thereby, privacy of information inside the hierarchies remains preserved. Similarly, in voltage control, as the local VAR resources are used to compensate for the reactive power imbalance, information about the entire system is not required to execute the sequential APPF.

\section{Simulation Case studies}\label{simulation_result}

We validate our APPF-based hierarchical frequency and voltage control algorithm using the 33-bus, 3-area power system model, shown in Fig. \ref{33_Bus_Case_Study}. Each area consists of standard IEEE 3-machine, 9-bus model integrated with 2 IBRs. The model parameters for the 9-bus system can be found in \cite{2018_M_A_Pai_Chow}. The SG dispatches and loads are modified to achieve non-zero tie-line flows between areas in pre-contingency steady-state. The dynamic model parameters of the IBRs are borrowed from \cite{2014_NREL_IBR_Model}. The maximum power rating of each IBR is 75.48 MW. The MVA base for whole system is 100 MVA. The series impedance and shunt admittance values of tie lines are $z_{tie} = 0.05 + j0.20$ p.u., $y_{tie} = j0.15$ p.u. respectively. We consider $w_1 = w_2 = 1$ for the following case-studies. As per the requirement from CPS implementation, total 7 PMUs are required per control area in Fig. \ref{33_Bus_Case_Study}. It is also assumed that each PMU channel sends 60 measurement samples per second.

\subsection{APPF steady-state performance evaluation for frequency control}
\begin{figure}[h]
	\begin{center}
		\includegraphics[width=\linewidth]{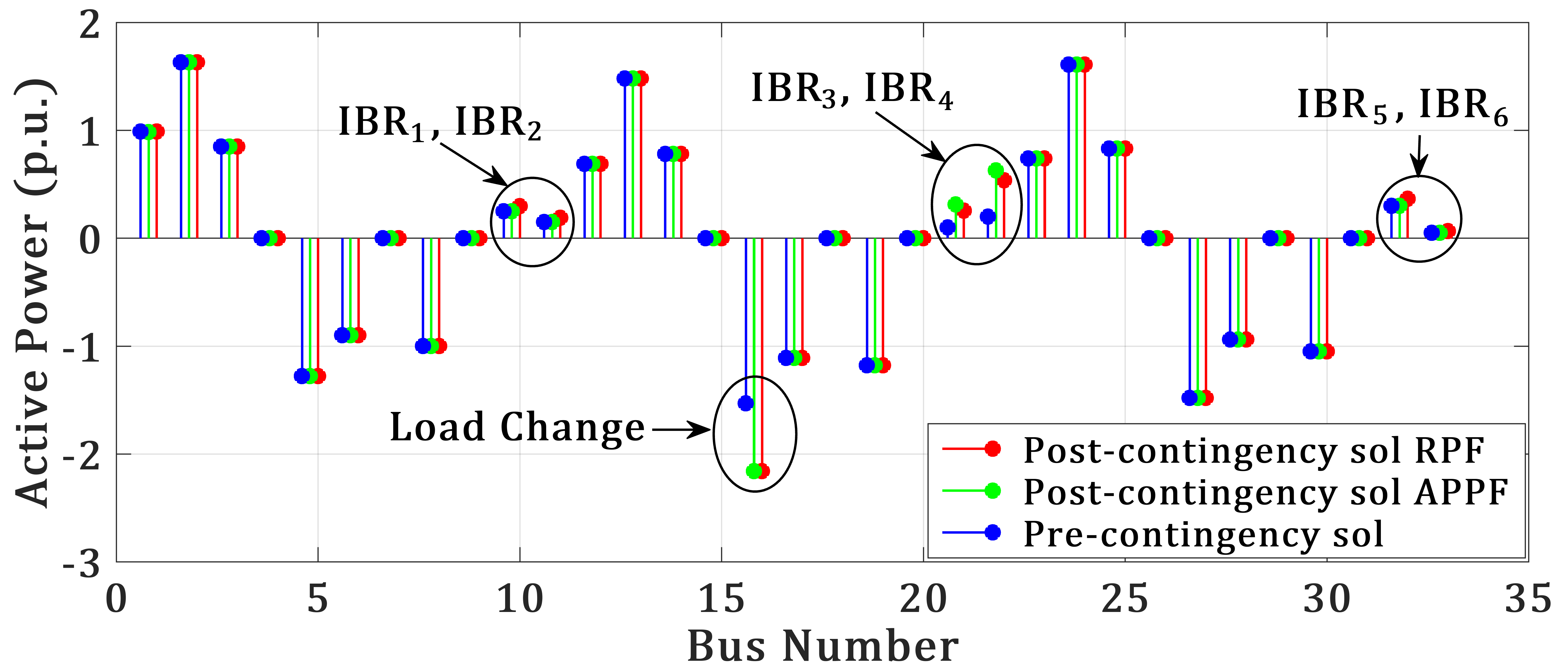}
		\vspace{-15pt}
		\caption{Active Power ($P$) injection in the buses for case 1}
		\label{APPF_case1_P}
	\end{center}
\end{figure}

\noindent \textbf{Case 1:} {\it Contingent area is self-sufficient} - The contingency is simulated by increasing the active power load at bus 16 by 63 MW which is 16.5\% increase with respect to the total active power load of the contingent area $A_1^{H_1}$. IBR$_3$ and IBR$_4$ in $A_1^{H_1}$ have sufficient cumulative capacity to compensate for this load change. Fig. \ref{APPF_case1_P} shows the post-contingency active power setpoints generated by APPF for the 33 buses. Note that the loads are indicated as negative power injections. The figure shows that the load change is fully compensated by increase in the active power output of the IBR$_3$, IBR$_4$ connected to buses 21 and 22, respectively.

\begin{figure}[H]
	\begin{center}
		\includegraphics[width=\linewidth]{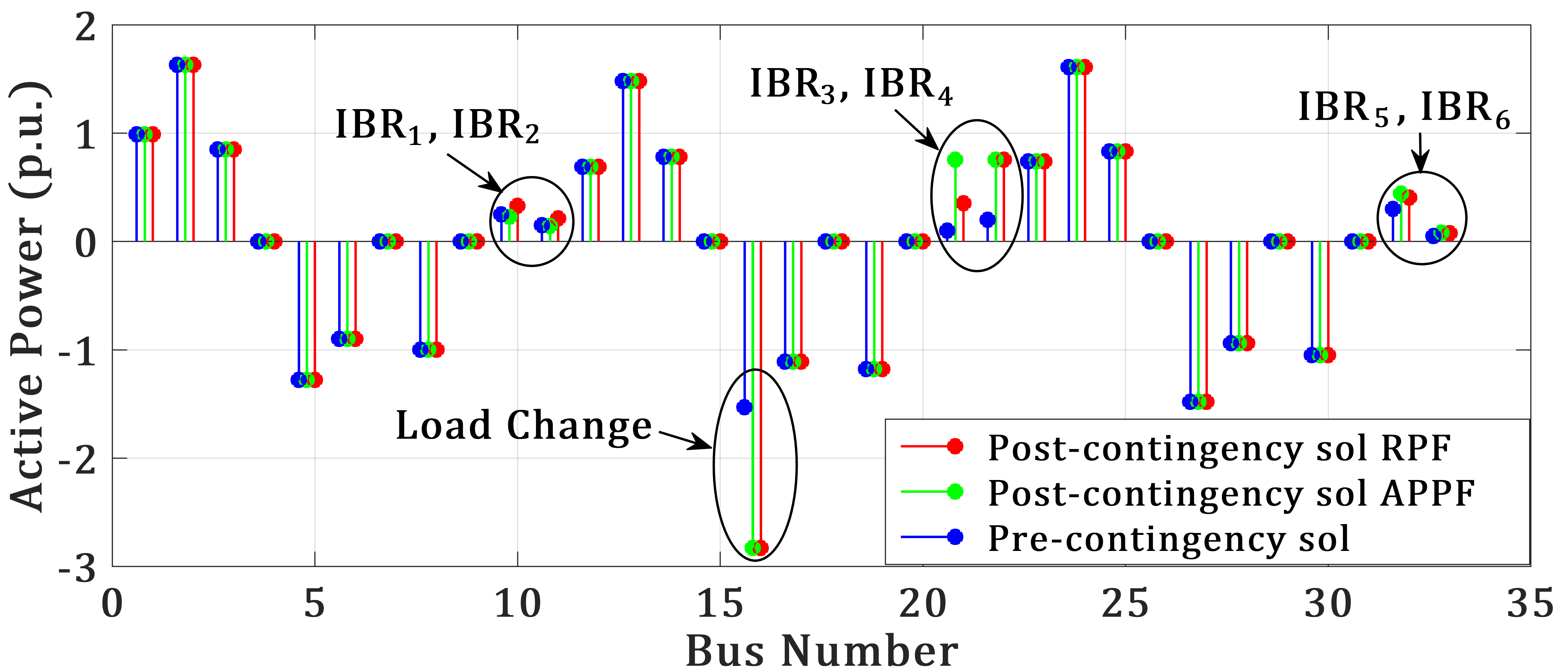}
		\vspace{-15pt}
		\caption{Active Power ($P$) injection in the buses for case 2}
		\label{APPF_case2_P}
	\end{center}
\end{figure}

\noindent \textbf{Case 2:} {\it Contingent area is self-deficient} - The load at bus 16 is now increased by 130 MW which is 34\% increase with respect to the total active power load of the contingent area. IBR$_3$ and IBR$_4$ are not sufficient to compensate for this change. Fig. \ref{APPF_case2_P} shows how IBR$_1$, IBR$_2$, IBR$_5$ and IBR$_6$ contribute the deficit power in this situation. Also note that in both figures \ref{APPF_case1_P} and \ref{APPF_case2_P}, the setpoints for the non-IBR buses (i.e., synchronous generator buses) do not change. The figures also reveal that the RPF solution for the IBR buses do not make full use of the available IBR capacity in the contingent area. The APPF solution, on the other hand, maximizes this utilization.
\vspace{-11.5pt}
\vspace{-25pt}
\vspace{-10pt}

\subsection{Steady-state performance evaluation for voltage control}

\begin{figure}[h]
	\begin{center}
		\includegraphics[width=\linewidth]{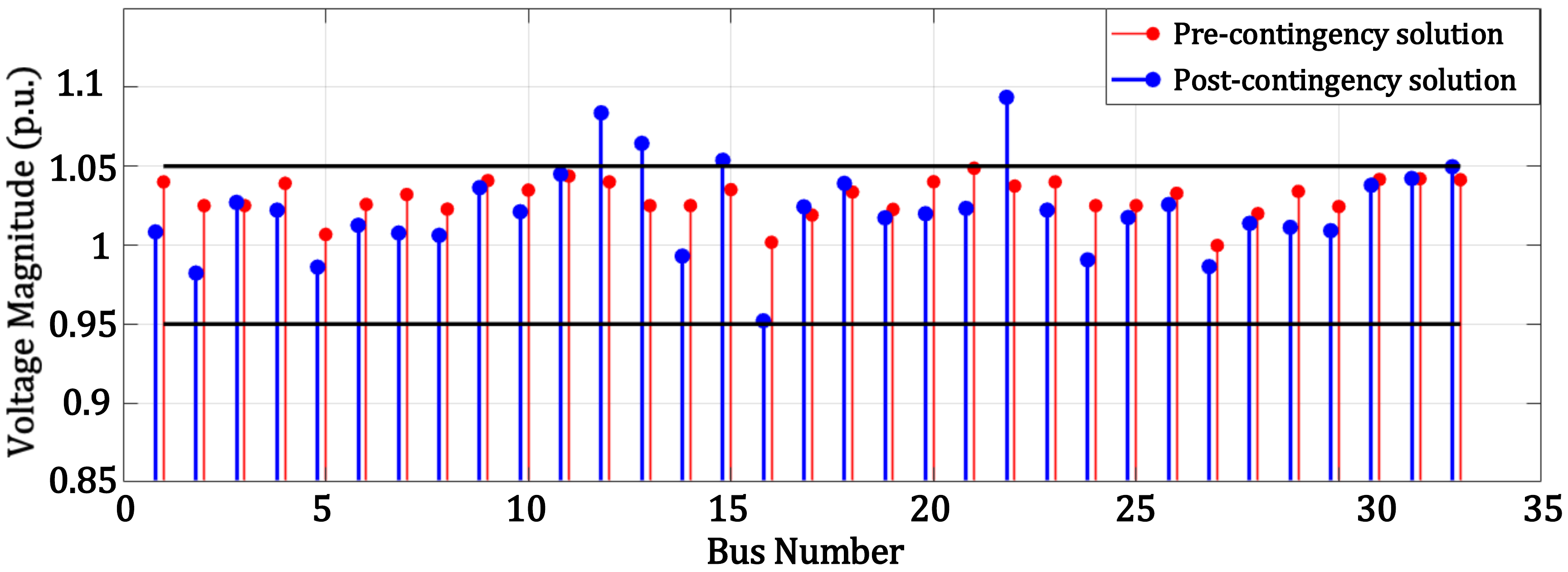}
		\vspace{-15pt}
		\caption{Bus voltages after solving voltage control optimization}
		\label{Volt_Control_ss}
	\end{center}
\end{figure}

We treat the 33 bus system as a whole for one area case study and utilize the hierarchies in generation resources to evaluate the proposed voltage control strategy. A reactive power change of 105 MVAR is simulated at load bus 16. This is 30.4\% increase with respect to the total reactive power load of the system. IBR$_1$, IBR$_2$, IBR$_3$, IBR$_4$, IBR$_5$ and IBR$_6$ change their reactive power setpoints to participate in the reactive power compensation. Also note that the synchronous generator buses adjust their voltage setpoints as shown in fig. \ref{Volt_Control_ss}. 
\vspace{-10pt}
\\
\vspace{-20pt}

The participation of SGs keep most of the bus voltages within the local band of 5\% around the nominal value. However, as per algorithm 1, we consider global bound for the voltages to be 10\% around the pre-contingency equilibrium. Therefore, voltage relaxation is allowed to the highly ranked IBR bus 22 nearest to the contingency to extract maximum reactive power from that IBR. Also, neighboring buses 12, 13, 15 are relaxed to their global voltage bound which is designed to be 1.1 per unit for the simulated system.

\subsection{Dynamic performance evaluation for frequency control}

\begin{figure}[h]
     \begin{center}
     \begin{subfigure}[h]{4cm}
         \begin{center}
         \includegraphics[width=\textwidth]{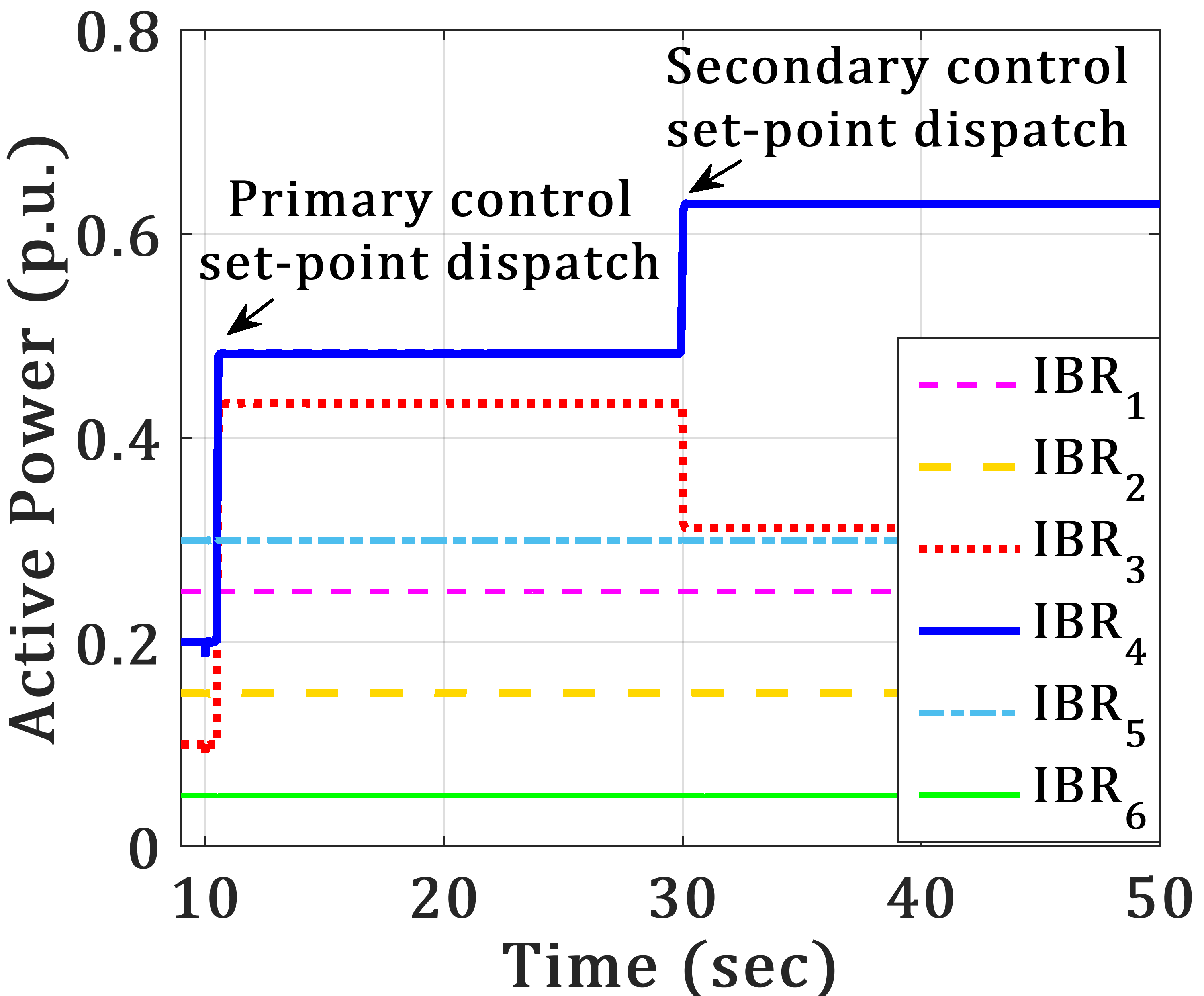}
         \vspace{-12.5pt}
         \caption{}
         \label{Case1_IBR_P_wo_droop_w_APPF}
         \end{center}
     \end{subfigure}
     \begin{subfigure}[h]{4cm}
        \begin{center}
         \includegraphics[width=\textwidth]{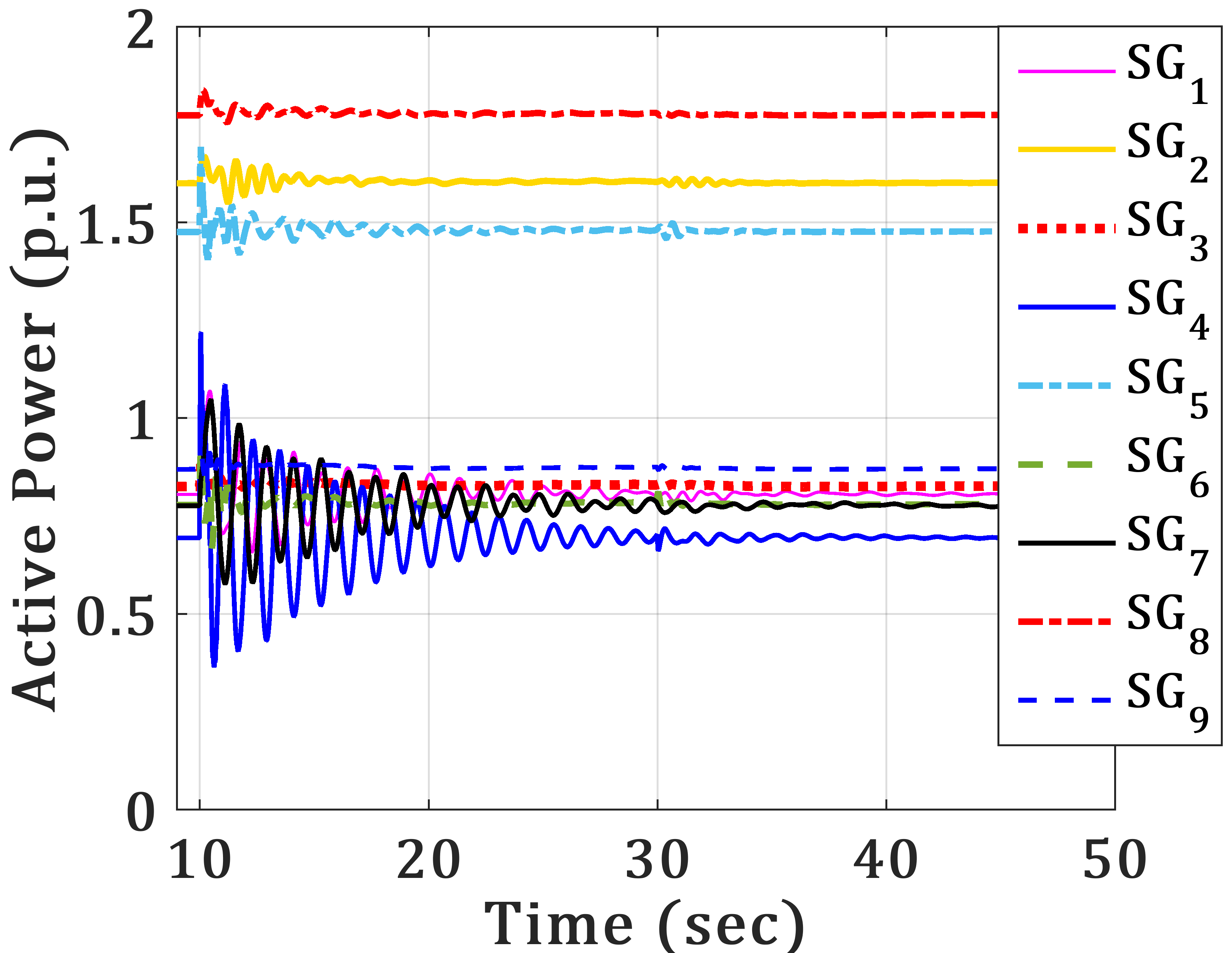}
         \vspace{-12.5pt}
         \caption{}
         \label{Case1_SG_P_wo_droop_w_APPF}
         \end{center}
     \end{subfigure}
     \vspace{-10pt}
        \caption{Active power output trajectories of (a) IBRs and (b) SGs with hierarchical update of IBR setpoints in case 1}
        \label{Case1_P_wo_droop_w_APPF}
    \end{center}
\end{figure}

\begin{figure}[h]
	\begin{center}
		\includegraphics[width=\linewidth]{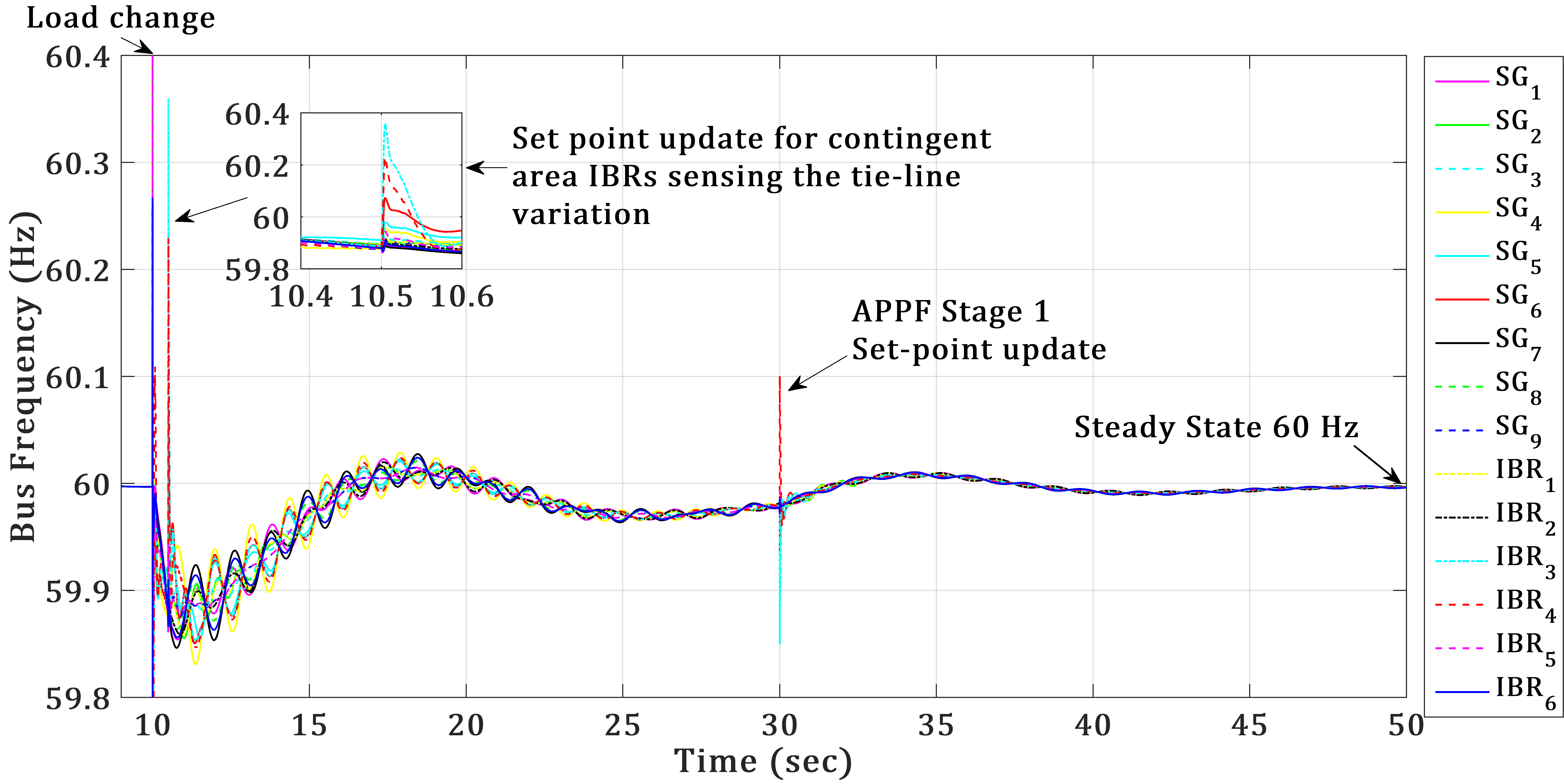}
		\vspace{-10pt}
		\caption{Bus frequency trajectories for case 1 with hierarchical updates of IBR$_3$ and IBR$_4$ active power setpoints}
		\label{Case1_Freq_wo_droop_w_APPF}
	\end{center}
\end{figure}

\begin{figure}[h]
	\begin{center}
		\includegraphics[width=\linewidth]{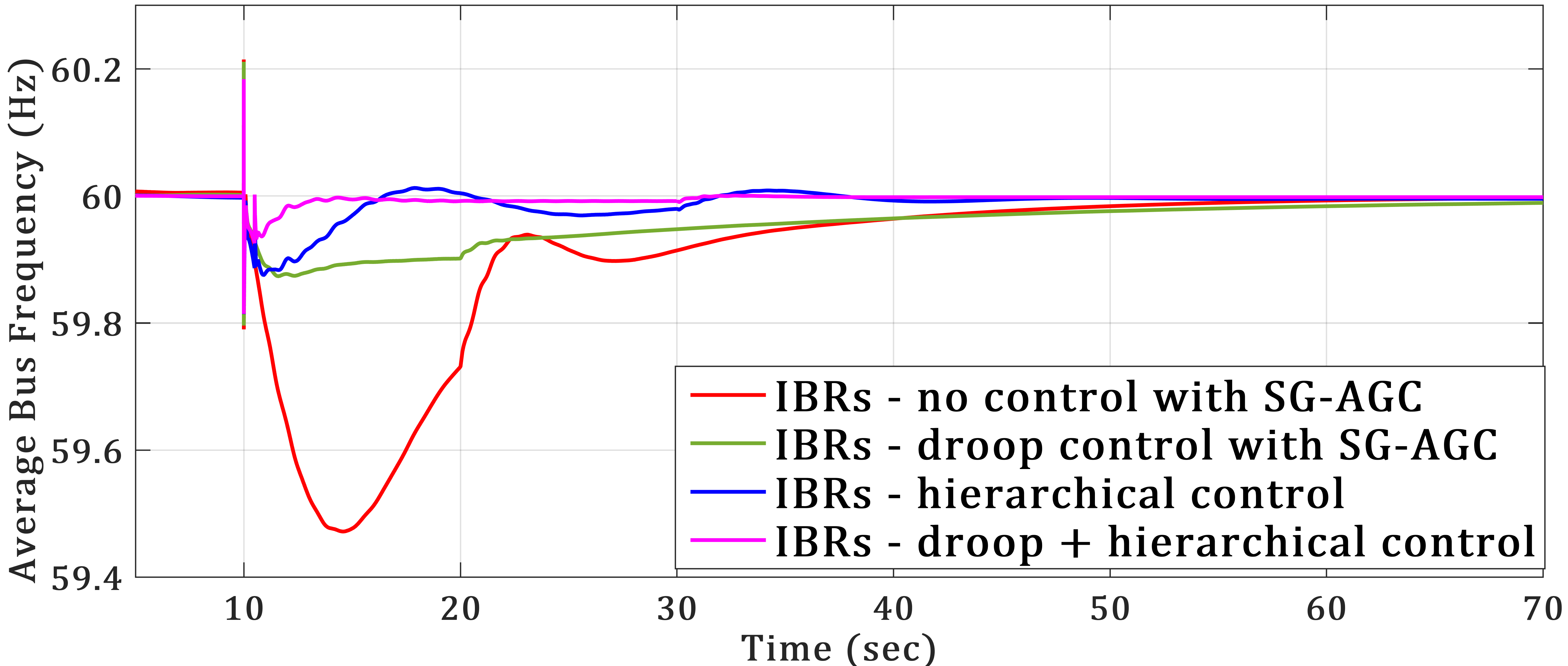}
		\vspace{-10pt}
		\caption{Comparison of different control scenarios for case 1}
		\label{Case1_scenario_comparison}
	\end{center}
\end{figure}

\noindent \textbf{Case 1:} {\it Contingent area is self-sufficient} - We simulate the active power load change at bus 16 by 63 MW at $t$ = 10 sec. The setpoints of IBRs in $A_1^{H_1}$ are updated using primary control dispatch at $t$ = 10.5 sec. The time constant for varying the IBR power output is assumed to be 0.01 sec following \cite{2012_WECC_Report}. Note that, the deviation in tie-line power flows give an approximate estimation of amount of contingency. Therefore, the state estimators of the contingent area may take time to precisely identify the location and the amount of contingency needed for solving the first stage of APPF. Let this time delay be 20 seconds. The active power setpoints of IBR$_3$ and IBR$_4$ are updated at $t$ = 30 sec according to the post-contingency APPF solution, listed in Fig. \ref{APPF_case1_P}. The active power output trajectories of IBRs are shown in Fig. \ref{Case1_IBR_P_wo_droop_w_APPF} which represents the fast response of IBRs after the dispatch of primary and secondary control setpoints. Following initial transients, SGs settle down to their pre-contingency outputs as IBRs are able to fully compensate for the active power imbalance through the proposed control strategy. The active power output trajectories of SGs are portrayed in Fig. \ref{Case1_SG_P_wo_droop_w_APPF}. The time response of the bus frequencies are shown in Fig. \ref{Case1_Freq_wo_droop_w_APPF}. As the IBRs of the contingent area have sufficient capacity to compensate the load change, only APPF stage-1 setpoints affect the frequency trajectories in this case. The APPF setpoints overwrite the primary control setpoints at $t$ = 30 sec after solving the local power flow for the $1^{st}$ hierarchy. Fig. \ref{Case1_scenario_comparison} compares four different scenarios for this case. The red curve shows the frequency response with no control applied to IBRs. The primary control of SGs act for the initial period till $t$ = 20 sec before AGC is triggered and restores the frequency to 60 Hz. The green curve shows the effect of adding droop control in IBRs. Both curves have low frequency nadir and take time till $t$ = 70 sec to reach 60 Hz due to the slower timescale of AGC operation \cite{2021_Duncan,2021_Nikos_AGC_timescale}. Our hierarchical control, in contrast, as shown by the blue and magenta curves successfully drives the frequency to 60 Hz, and that too much faster than AGC and with better transient performance. The hierarchical control in presence of droop in this case shows the best dynamic performance.
\vspace{-6pt}
\vspace{-25pt}

\noindent \textbf{Case  2:} \textit{Contingent area is self-deficient} - We next simulate the active power load change of 130 MW at bus 16 at $t$ = 10 sec. Inter-area communication delays are incorporated for the simulation of self-deficient case. An open communication network may have transmission delay changing within [0.15 sec, 2 sec]. We consider 0.25 sec as the transmission delay margin for the inter-area communication. 
\begin{figure*}[!b]
	\begin{center}
		\includegraphics[width=\linewidth]{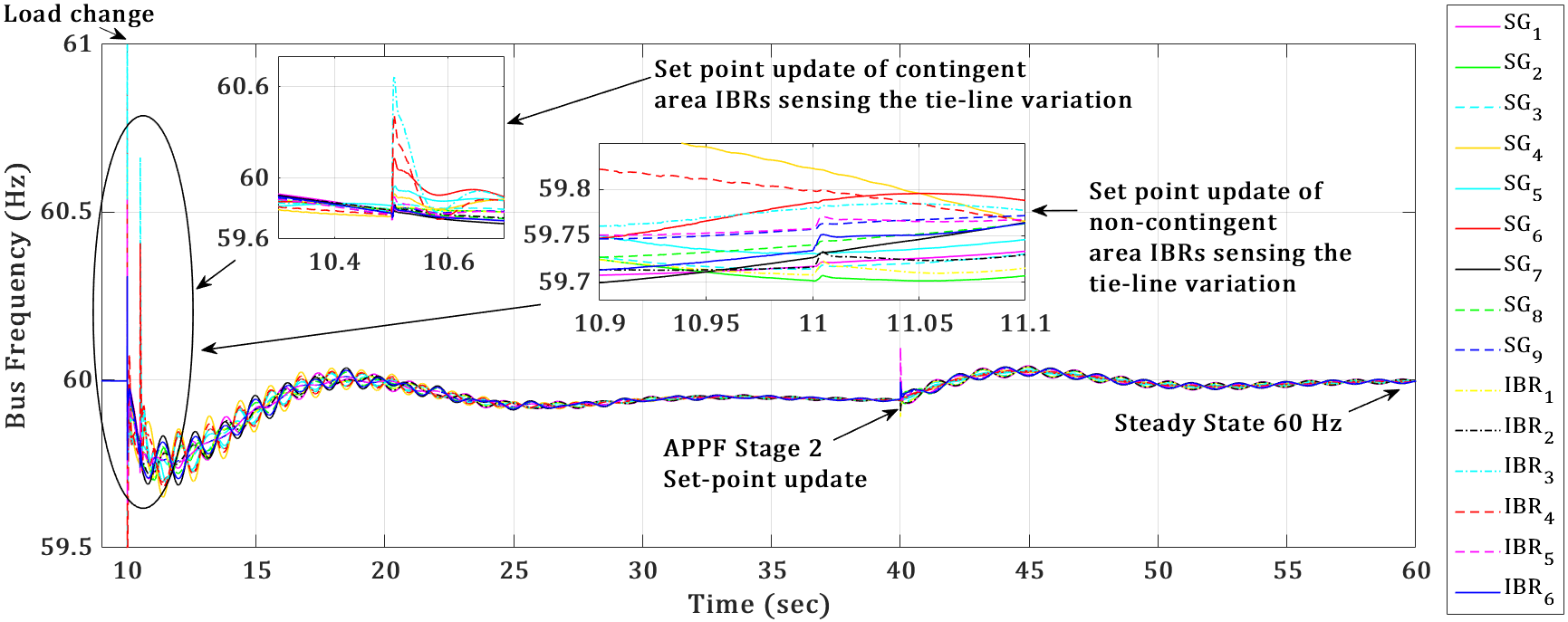}
		\vspace{-15pt}
		\caption{Frequency trajectories for case 2 with hierarchical updates of active power setpoints of IBR$_1$ through IBR$_6$}
		\label{Case2_Freq_wo_droop_w_APPF}
	\end{center}
\end{figure*}

\begin{figure}[h]
     \centering
     \begin{subfigure}[h]{4cm}
         \centering
         \includegraphics[width=\textwidth]{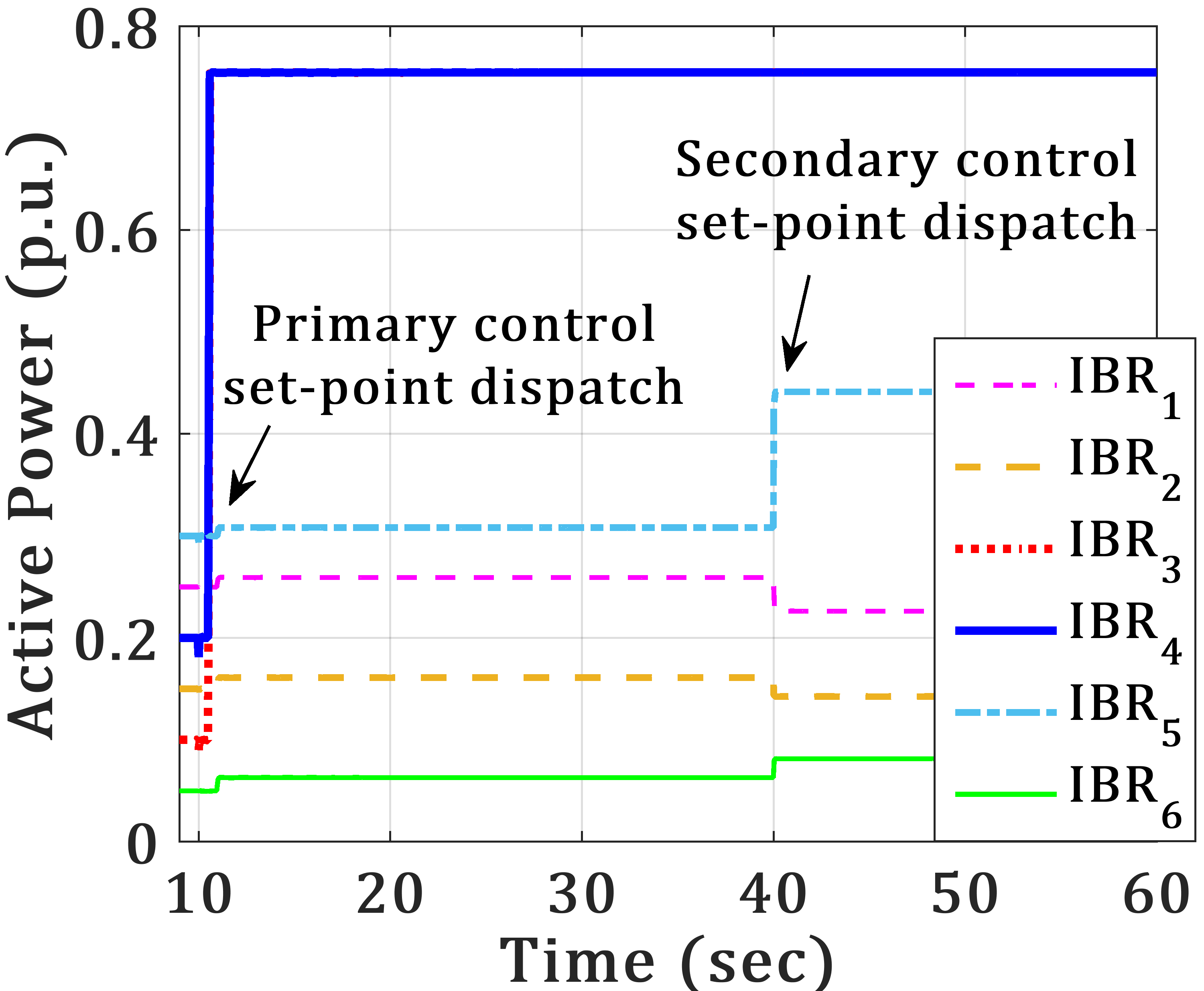}
         \vspace{-12.5pt}
         \caption{}
         \label{Case2_IBR_P_wo_droop_w_APPF}
     \end{subfigure}
     \begin{subfigure}[h]{4cm}
         \centering
         \includegraphics[width=\textwidth]{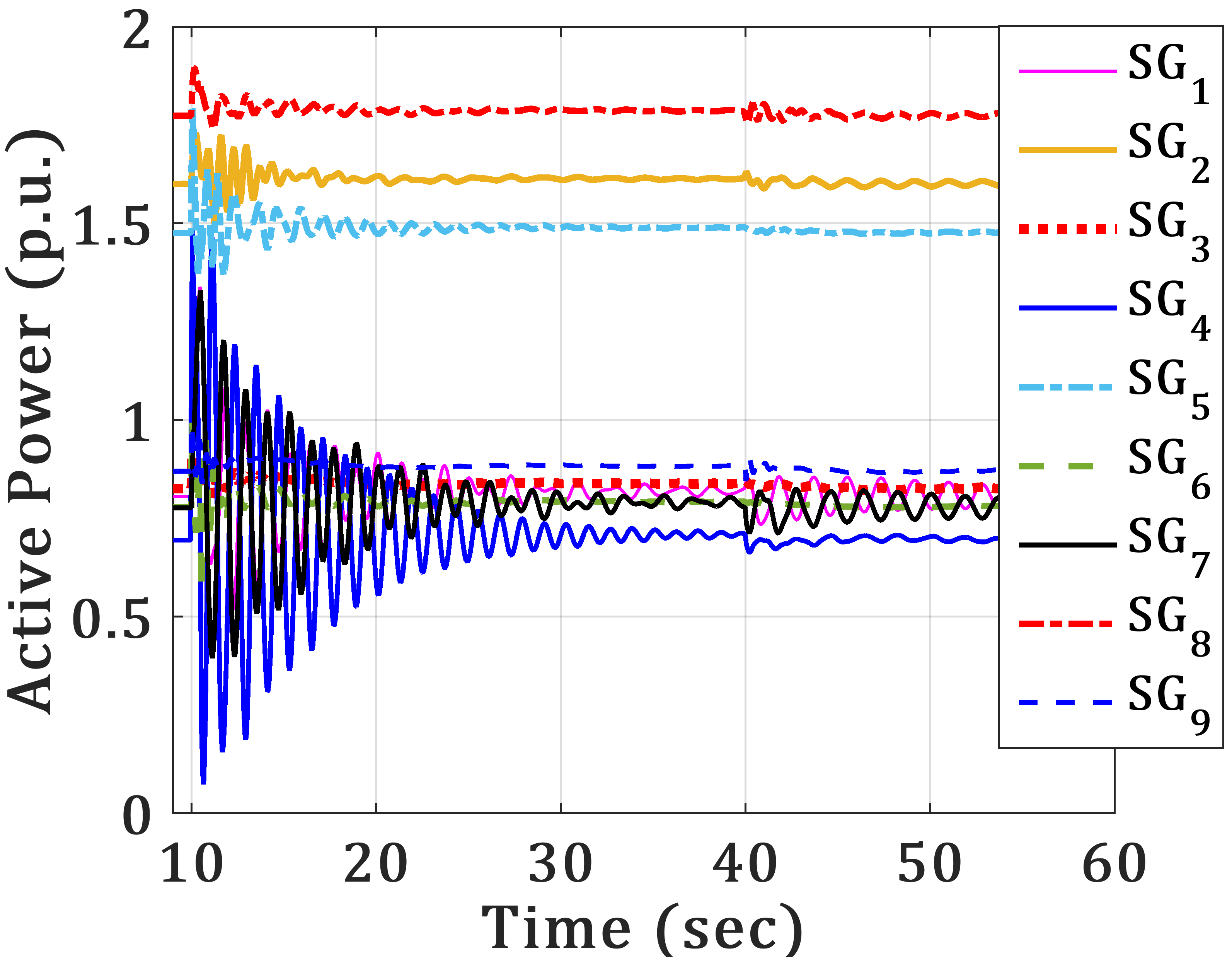}
         \vspace{-12.5pt}
         \caption{}
         \label{Case2_SG_P_wo_droop_w_APPF}
     \end{subfigure}
     \vspace{-10pt}
        \caption{Active power output trajectories of (a) IBRs and (b) SGs with hierarchical update of IBR setpoints in case 2}
        \label{Case2_P_wo_droop_w_APPF}
\end{figure}

\begin{figure}[h]
	\begin{center}
		\includegraphics[width=\linewidth]{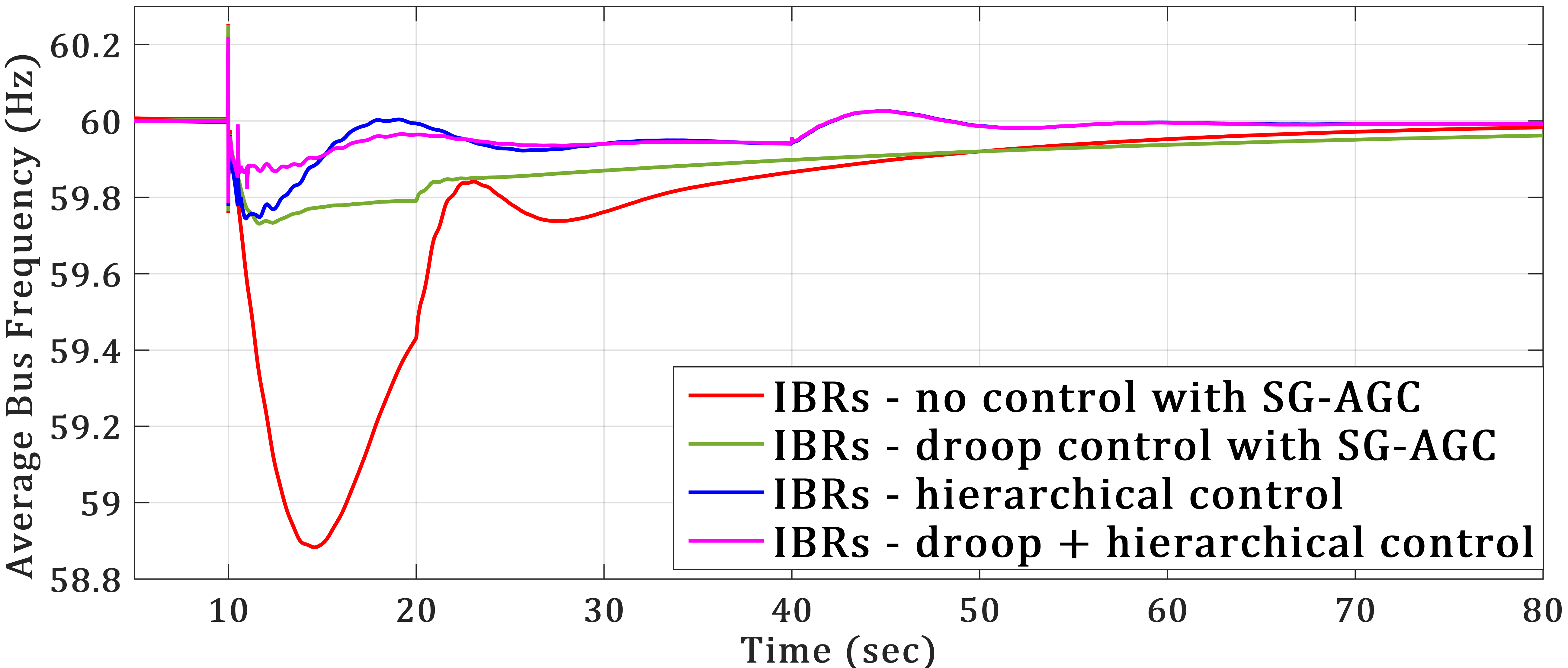}
		\vspace{-10pt}
		\caption{Comparison of different control scenarios for case 2}
		\label{Case2_scenario_comparison}
	\end{center}
\end{figure}

The setpoints of IBRs in both contingent and non-contingent areas are updated from tie-line flow deviations following their available headrooms at $t$ = 10.5 sec and $t=11$ sec respectively. The set points of IBR$_1$ through IBR$_6$ are thereafter updated using the post-contingency APPF solutions from Fig. \ref{APPF_case2_P}. The time response of the bus frequencies are shown in Fig.\ref{Case2_Freq_wo_droop_w_APPF}. An important observation is that the updates of the setpoints from APPF stage 1 in this case do not have any effect on the dynamics of the frequency. This is because the primary control already uses the maximum power available from IBR$_3$ and IBR$_4$ as shown in Fig. \ref{Case2_IBR_P_wo_droop_w_APPF}. The only information provided by APPF stage 1 is the change in the tie-line power required from the other areas. APPF stage 2 updates the primary setpoints at $t$ = 40 sec after solving the power flow for the second hierarchy. Fig. \ref{Case2_IBR_P_wo_droop_w_APPF} depicts the fast response of IBRs follwoing the setpoint dispatch, whereas the active power output trajectories of SGs are portrayed in Fig. \ref{Case2_SG_P_wo_droop_w_APPF}. After initial transients, SGs settle down to their pre-contingency outputs which shows the sole participation of IBRs to control the frequency in a faster time-scale without engaging conventional generators. Please note that APPF does not influence the power oscillation damping of SGs as reflected in Fig. \ref{Case1_SG_P_wo_droop_w_APPF} and \ref{Case2_SG_P_wo_droop_w_APPF}. It is executed in time sensitive way to reduce the settling time of frequency convergence to 60 Hz. Fig. \ref{Case2_scenario_comparison} shows the comparison of four different scenarios. Similar to Fig. \ref{Case1_scenario_comparison}, it is observed that our proposed hierarchical control outperforms both SG-based and droop-based primary and AGC controls in terms of dynamic performance of the frequencies.\\
\vspace{-15pt}

\subsection{Dynamic performance evaluation for generator trip}\label{dynamic_freq_cont_gen_trip}

\begin{figure}[h]
     \centering
     \begin{subfigure}[h]{4cm}
         \centering
         \includegraphics[width=\textwidth]{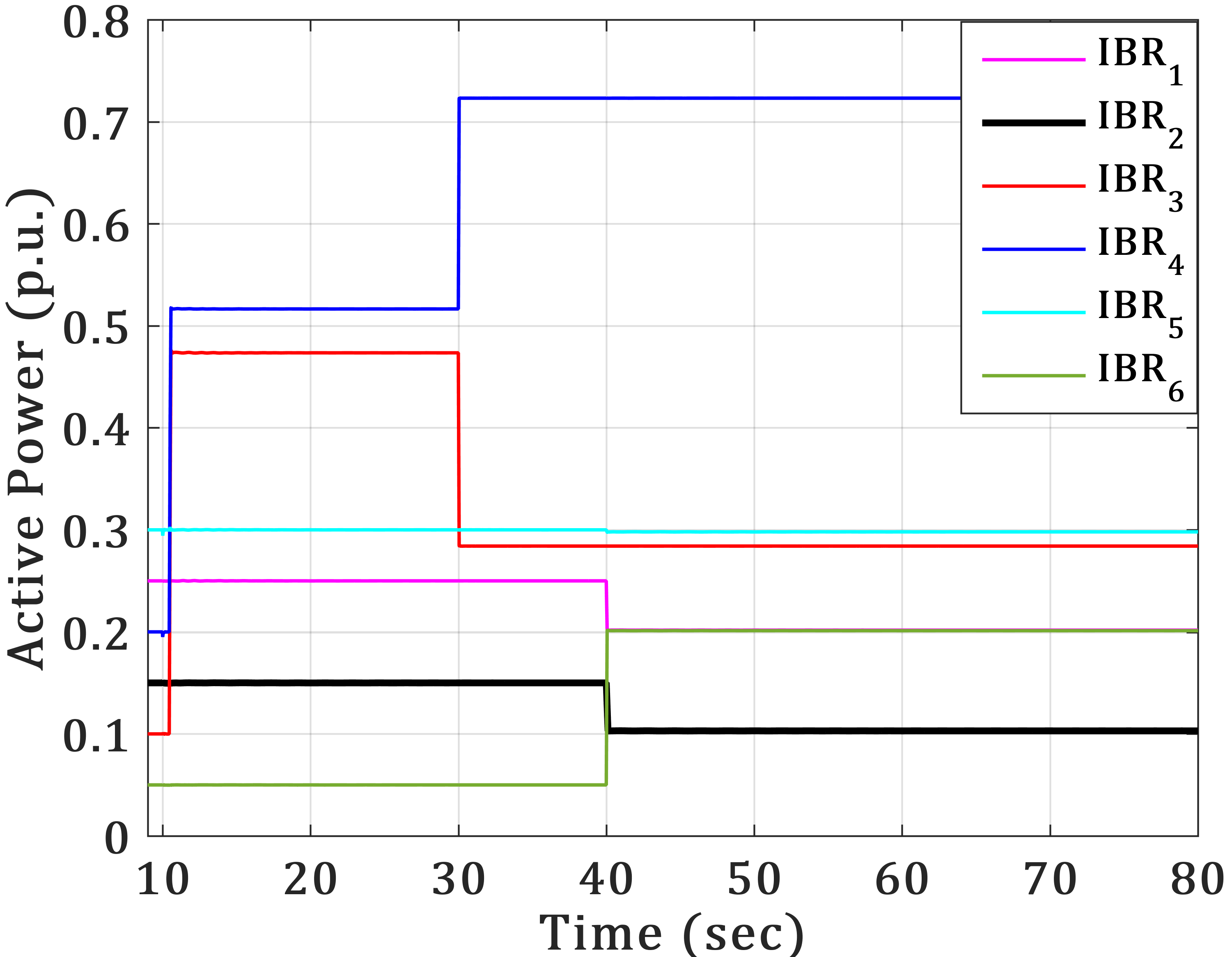}
         \vspace{-10pt}
         \caption{}
         \label{Plot_IBR_P_gen_trip}
     \end{subfigure}
     \begin{subfigure}[h]{4cm}
         \centering
         \includegraphics[width=\textwidth]{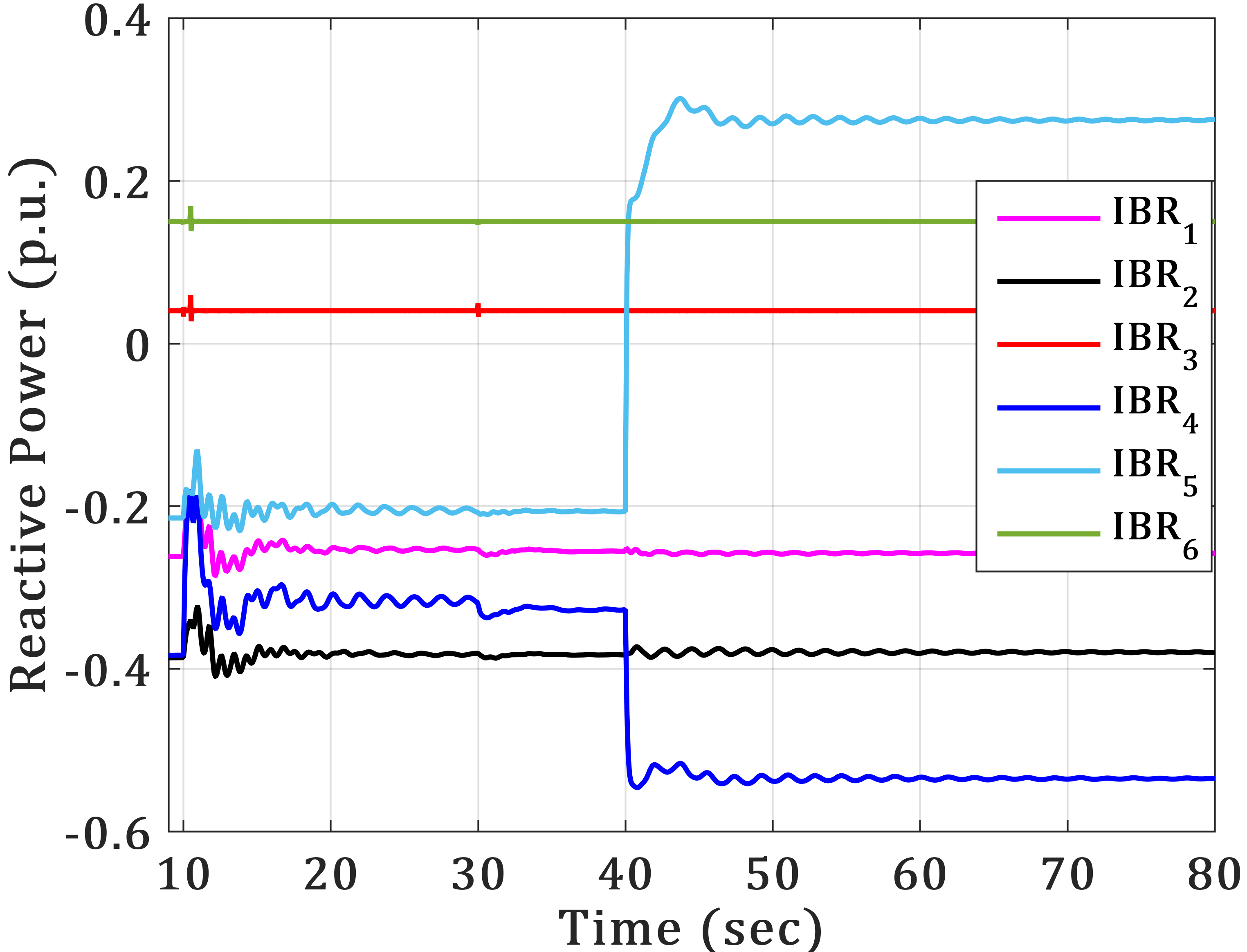}
         \vspace{-10pt}
         \caption{}
         \label{Plot_IBR_Q_gen_trip}
     \end{subfigure}
          \vspace{-8pt}
        \caption{(a) Active power outputs of IBRs, (b) Reactive power outputs of IBRs for generator $SG_4$ trip contingency}
        \label{Plot_IBR_PQ_gen_trip}
\end{figure}

\begin{figure}[h]
     \centering
     \begin{subfigure}[h]{4cm}
         \centering
         \includegraphics[width=\textwidth]{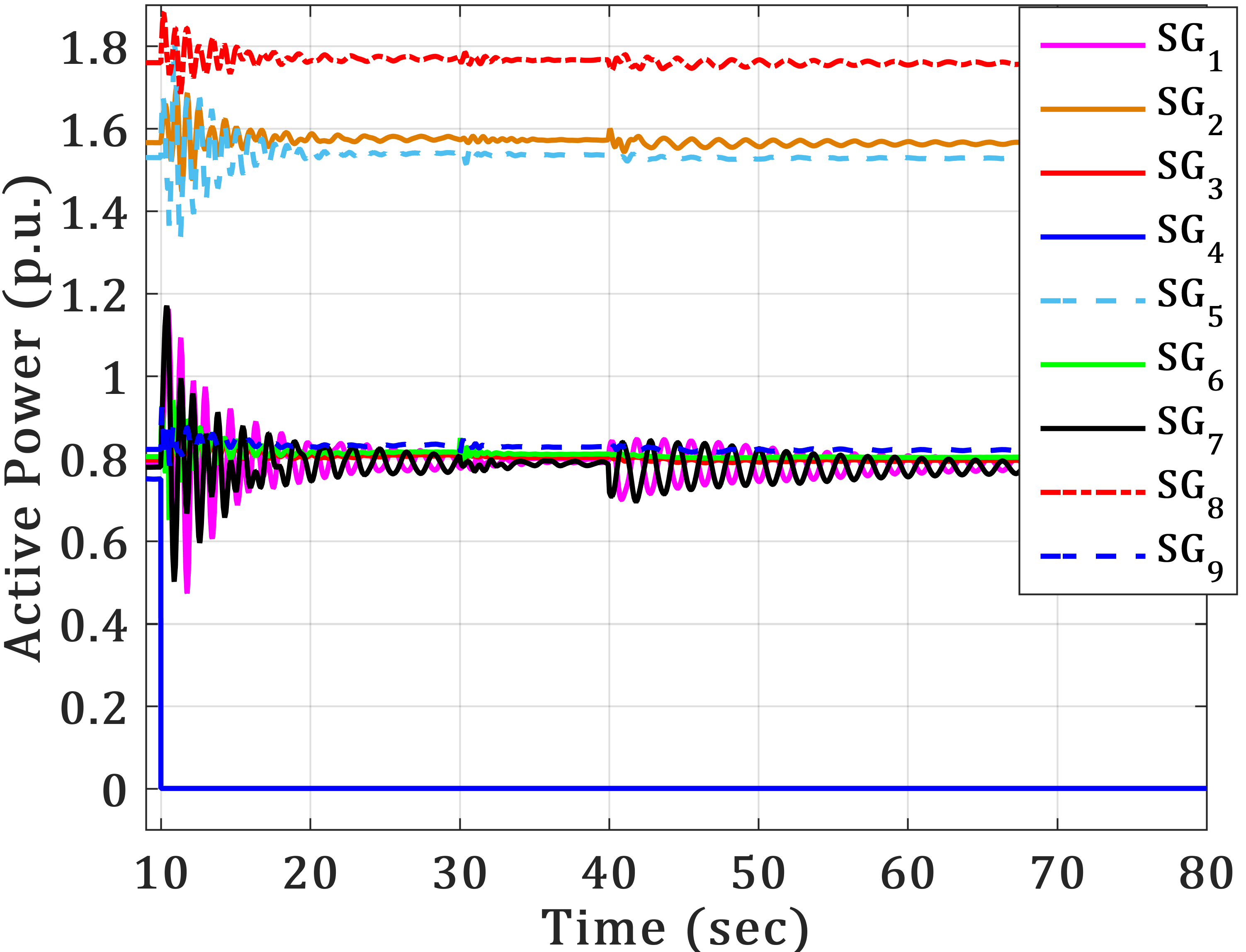}
         \vspace{-10pt}
         \caption{}
         \label{Plot_SG_P_gen_trip}
     \end{subfigure}
     \begin{subfigure}[h]{4cm}
         \centering
         \includegraphics[width=\textwidth]{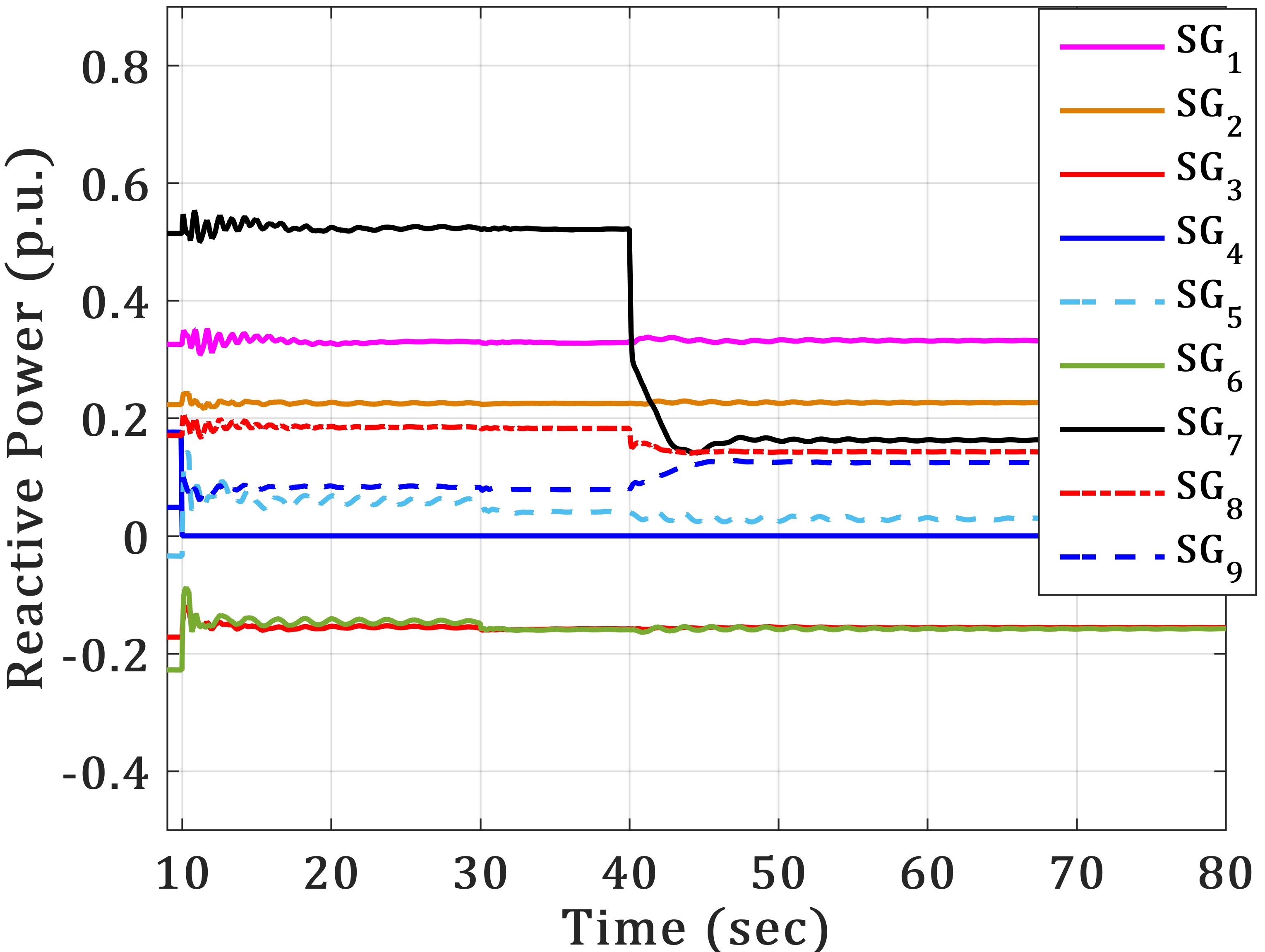}
         \vspace{-10pt}
         \caption{}
         \label{Plot_SG_Q_gen_trip}
     \end{subfigure}
     \vspace{-8pt}
        \caption{(a) Active power outputs of SGs, (b) Reactive power outputs of SGs for generator $SG_4$ trip contingency}
        \label{Plot_SG_PQ_gen_trip}
\end{figure}

\begin{figure}[h!]
	\begin{center}
		\includegraphics[width=\linewidth]{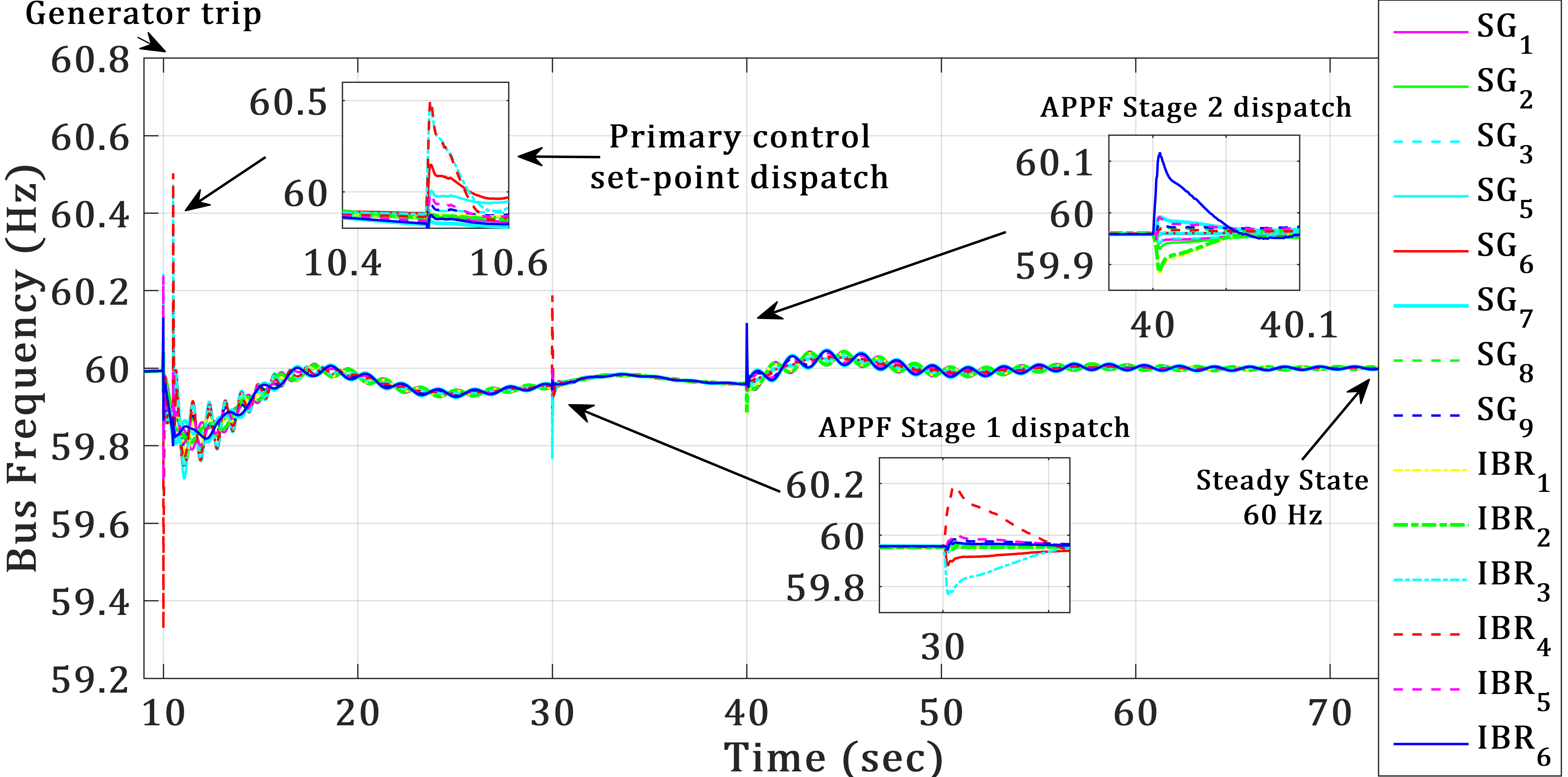}
		\vspace{-10pt}
		\caption{Bus frequencies for generator $SG_4$ trip contingency}
		\label{Freq_w_APPF_gen_trip}
	\end{center}
\end{figure}

Conventional generators are typically used for active power commitment with no fidelity to reactive power support in regulation market. Therefore, generator trips will be addressed by hierarchical frequency control. At $t$ = 10 sec, we simulate the tripping of $SG_4$ which was delivering 69 MW in pre-contingency condition. This change is equivalent to 21.2\% decrease with respect to the pre-contingency total active power generation of area $A_1^{H_1}$. Fig. \ref{Plot_IBR_P_gen_trip} shows the variation in active\\
\vspace{-15pt}

\noindent power outputs of IBRs with hierarchical dispatch of primary control setpoints, APPF stage 1 and stage 2 at 10.5 sec, 30 sec and 40 sec respectively. $IBR_3$ and $IBR_4$ have sufficient headroom to compensate for the estimated active power imbalance. However, to compensate for the exact change, secondary control executes both stages of APPF. To satisfy all the system operational constraints, APPF stage 1 is only able to utilize $IBR_4$ fully and $IBR_3$ partially. In absense of $SG_4$, the IBRs work in grid forming mode for the residual reactive power support as shown in Fig. \ref{Plot_IBR_Q_gen_trip}. Fig. \ref{Plot_SG_Q_gen_trip} shows that the other SGs also contribute to reactive power variation. However, there is no active power contribution from SGs in post-contingency steady state as portrayed in Fig. \ref{Plot_SG_P_gen_trip}. Fig. \ref{Freq_w_APPF_gen_trip} displays the time response of bus frequencies which get restored to 60 Hz after the hierarchical setpoint dispatch is executed.\\
\vspace{-15pt}
\subsection{Dynamic performance evaluation for voltage control}\label{dynamic_volt_cont_sim}
\begin{figure}[h]
     \centering
     \begin{subfigure}[h]{4cm}
         \centering
         \includegraphics[width=\textwidth]{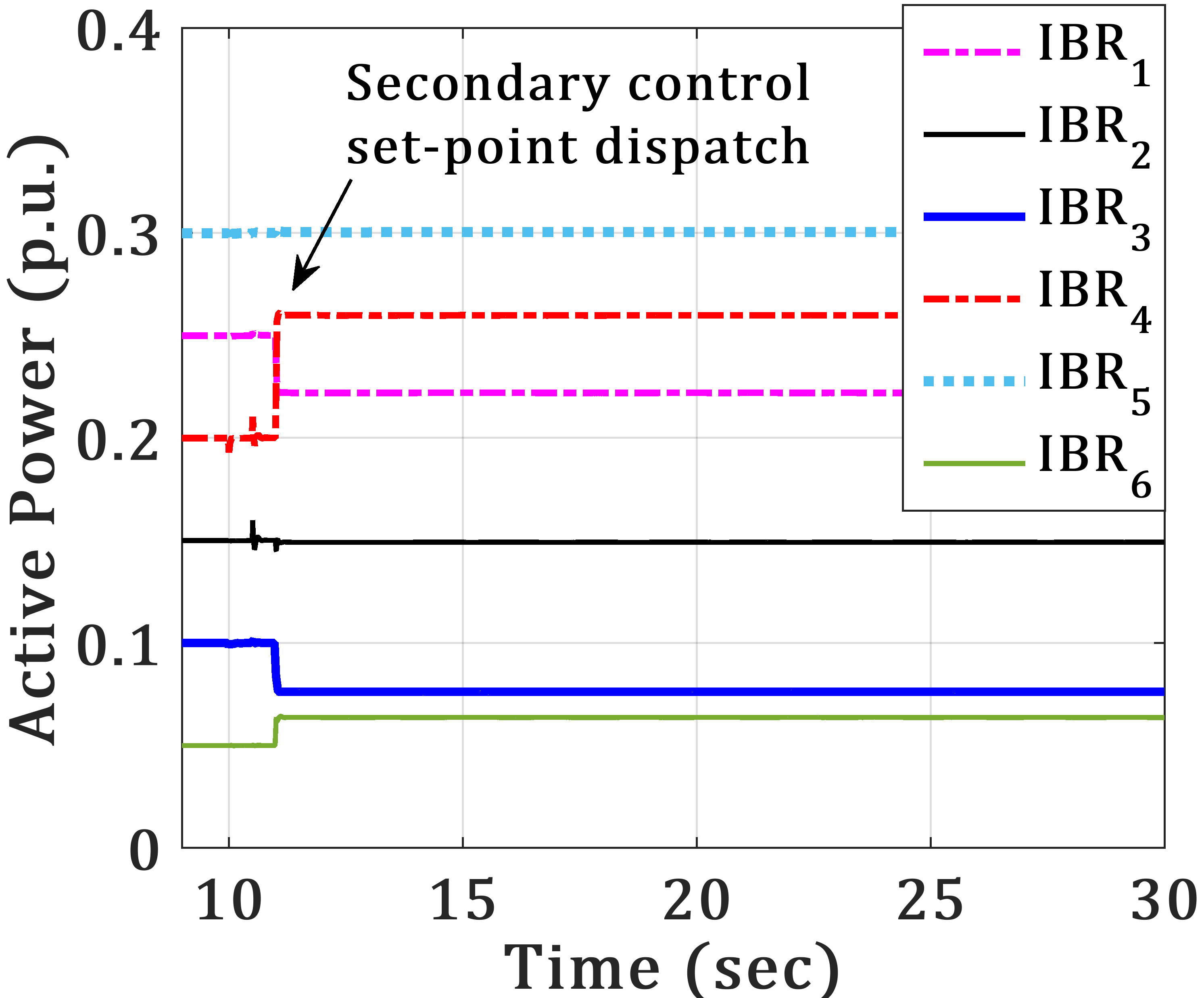}
         \vspace{-10pt}
         \caption{}
         \label{volt_IBR_P_wo_droop_w_APPF}
     \end{subfigure}
     \begin{subfigure}[h]{4cm}
         \centering
         \includegraphics[width=\textwidth]{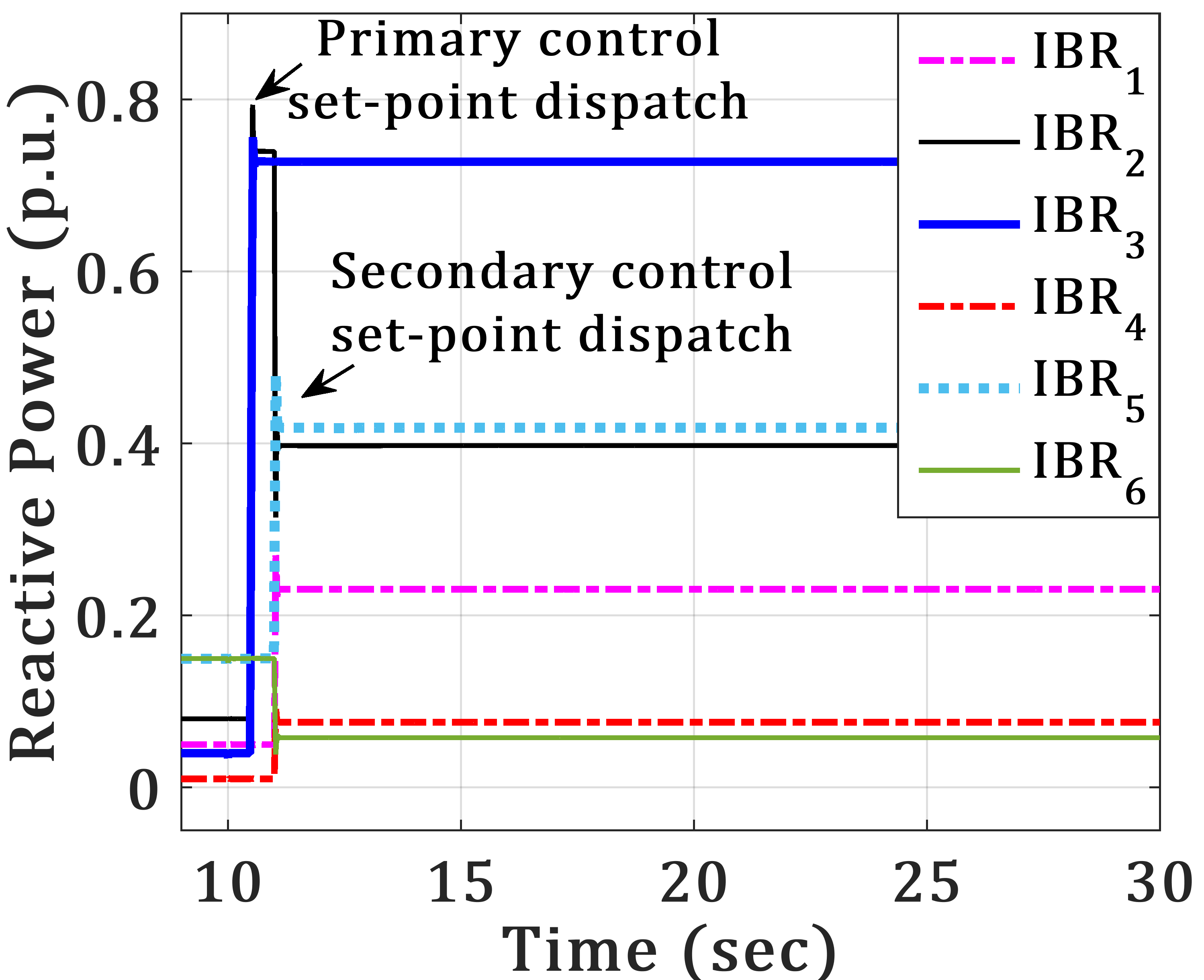}
         \vspace{-10pt}
         \caption{}
         \label{volt_IBR_Q_wo_droop_w_APPF}
     \end{subfigure}
          \vspace{-8pt}
        \caption{(a) Active power outputs of IBRs, (b) Reactive power outputs of IBRs with hierarchical update of setpoints}
        \label{volt_PQ_SG_IBR_wo_droop_w_APPF}
\end{figure}

\begin{figure}[h]
     \centering
     \begin{subfigure}[h]{4cm}
         \centering
         \includegraphics[width=\textwidth]{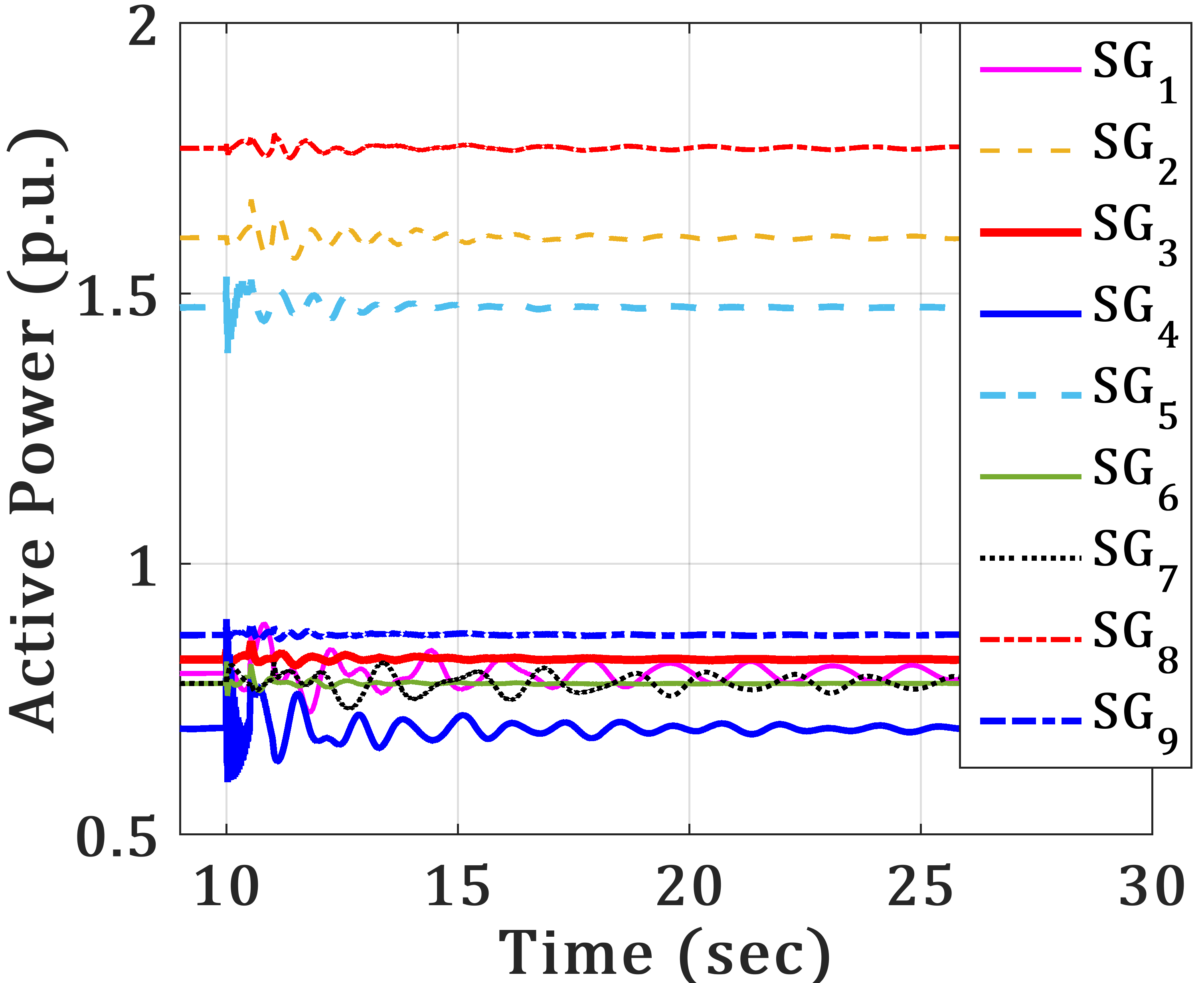}
         \vspace{-10pt}
         \caption{}
         \label{volt_SG_P_wo_droop_w_APPF}
     \end{subfigure}
     \begin{subfigure}[h]{4cm}
         \centering
         \includegraphics[width=\textwidth]{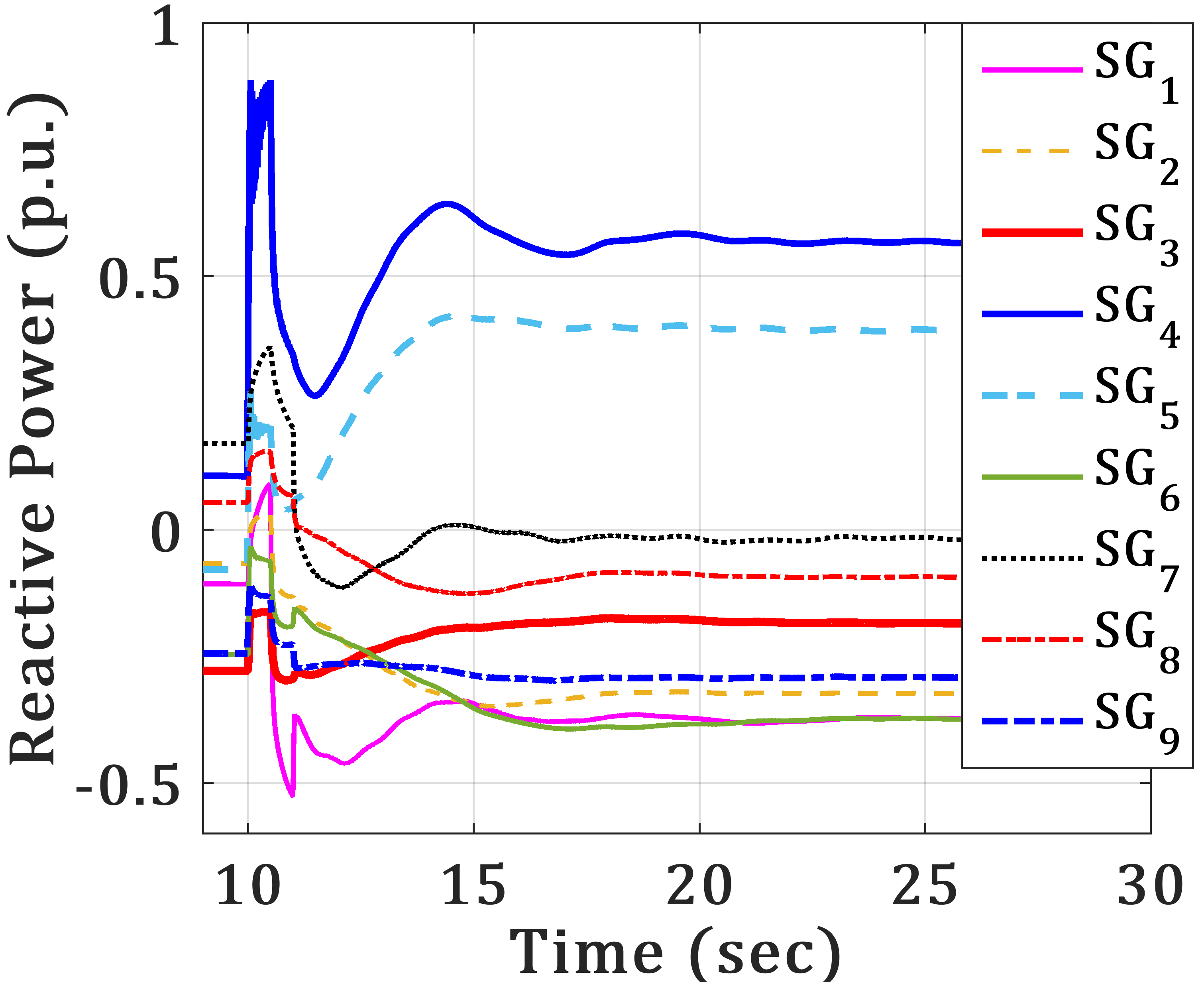}
         \vspace{-10pt}
         \caption{}
         \label{volt_SG_Q_wo_droop_w_APPF}
     \end{subfigure}
     \vspace{-8pt}
        \caption{(a) Active power outputs of SGs, (b) Reactive power outputs of SGs with hierarchical update of setpoints}
        \label{volt_PQ_SG_IBR_wo_droop_w_APPF}
\end{figure}

We simulate the reactive power load change at bus 16 by 105 MVAR at $t$ = 10 sec. The setpoints of IBRs are updated using primary control dispatch at $t$ = 10.5 sec. We consider the time delay to solve the secondary optimization as 0.5 second. The reactive power setpoints of IBRs are updated at $t$ = 11 sec according to the solution of optimization. The fast response of IBRs through active and reactive power setpoint dispatch is shown in Fig. \ref{volt_IBR_P_wo_droop_w_APPF} - \ref{volt_IBR_Q_wo_droop_w_APPF}. It is to be noted that active power outputs of IBRs are also adjusted by secondary optimization to compensate for the extra active power losses in the network while addressing the reactive power imbalance. The time response of the active and reactive power outputs of SG buses are represented in Fig. \ref{volt_SG_P_wo_droop_w_APPF} - \ref{volt_SG_Q_wo_droop_w_APPF}. Even though SGs\\

\begin{figure*}[!b]
	\begin{center}
		\includegraphics[width=\textwidth]{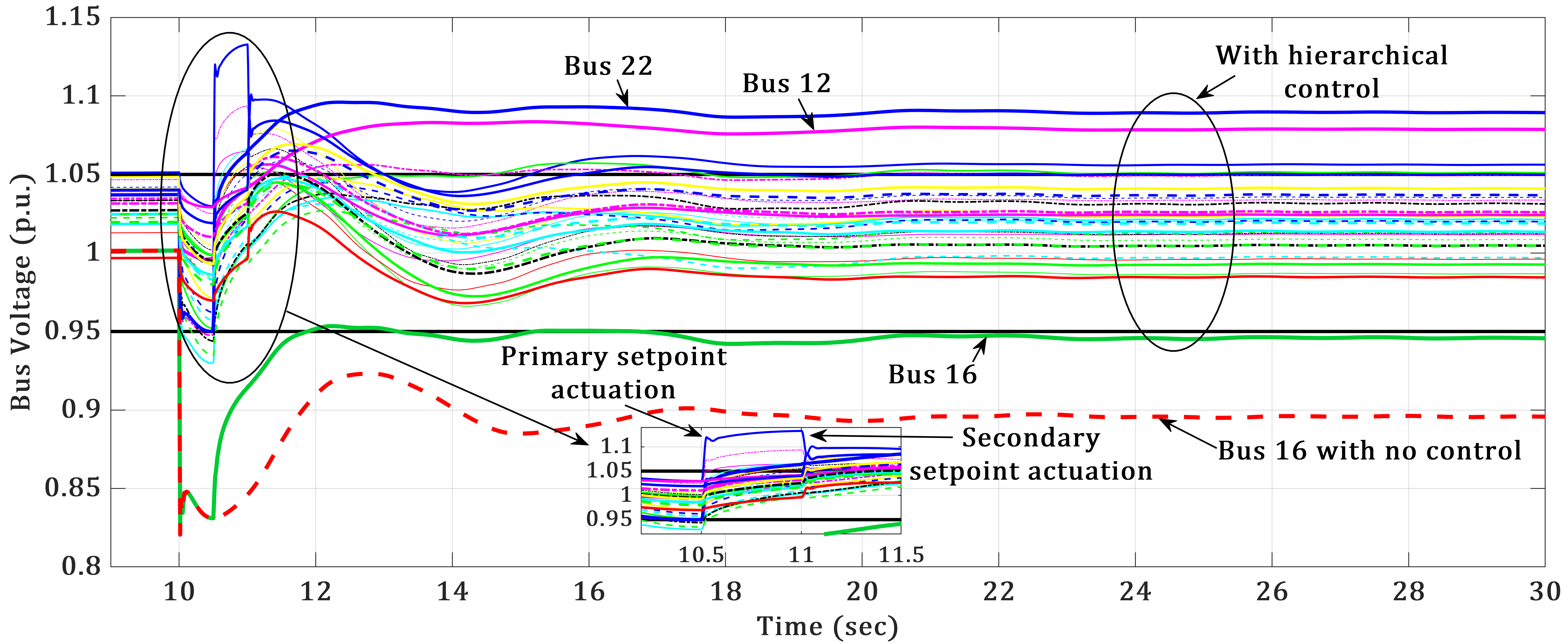}
		\caption{Voltage trajectories for APPF vs non-APPF}
		\label{Volt_Control_dynamics}
	\end{center}
\end{figure*}

\noindent participate in the reactive power compensation, their active power outputs settle down to pre-contingency outputs after the AVR setpoints are dispatched from secondary optimization. The bus voltage trajectories are shown in Fig. \ref{Volt_Control_dynamics}. From dynamic simulation, it is clear that most of the voltages are within the local band [0.95, 1.05] p.u. with the proposed hierarchical control. However, to maximally utilize IBR$_4$ which is closest to the contingent bus, we assign the global voltage bound to 10\% around the pre-contingency equilibrium to provide relaxation to the voltage when high reactive power injection happens at that particular IBR bus 22. The choice of the global voltage bound depends on the system operator. Its sole purpose is to attain a feasible solution for the secondary optimization while utilizing the highly ranked IBRs maximally. Therefore, in Fig. \ref{Volt_Control_dynamics} voltages of buses 12 and 22 settle to their respective post-contingency steady-states that are beyond the local bound of [0.95, 1.05] p.u., but within the global limit of 1.1 p.u. as explained in Fig. \ref{Volt_Control_ss}.\\
\vspace{-15pt}

\vspace{-15pt}

\begin{figure}[h]
     \centering
     \begin{subfigure}[h]{4cm}
         \centering
         \includegraphics[width=\textwidth]{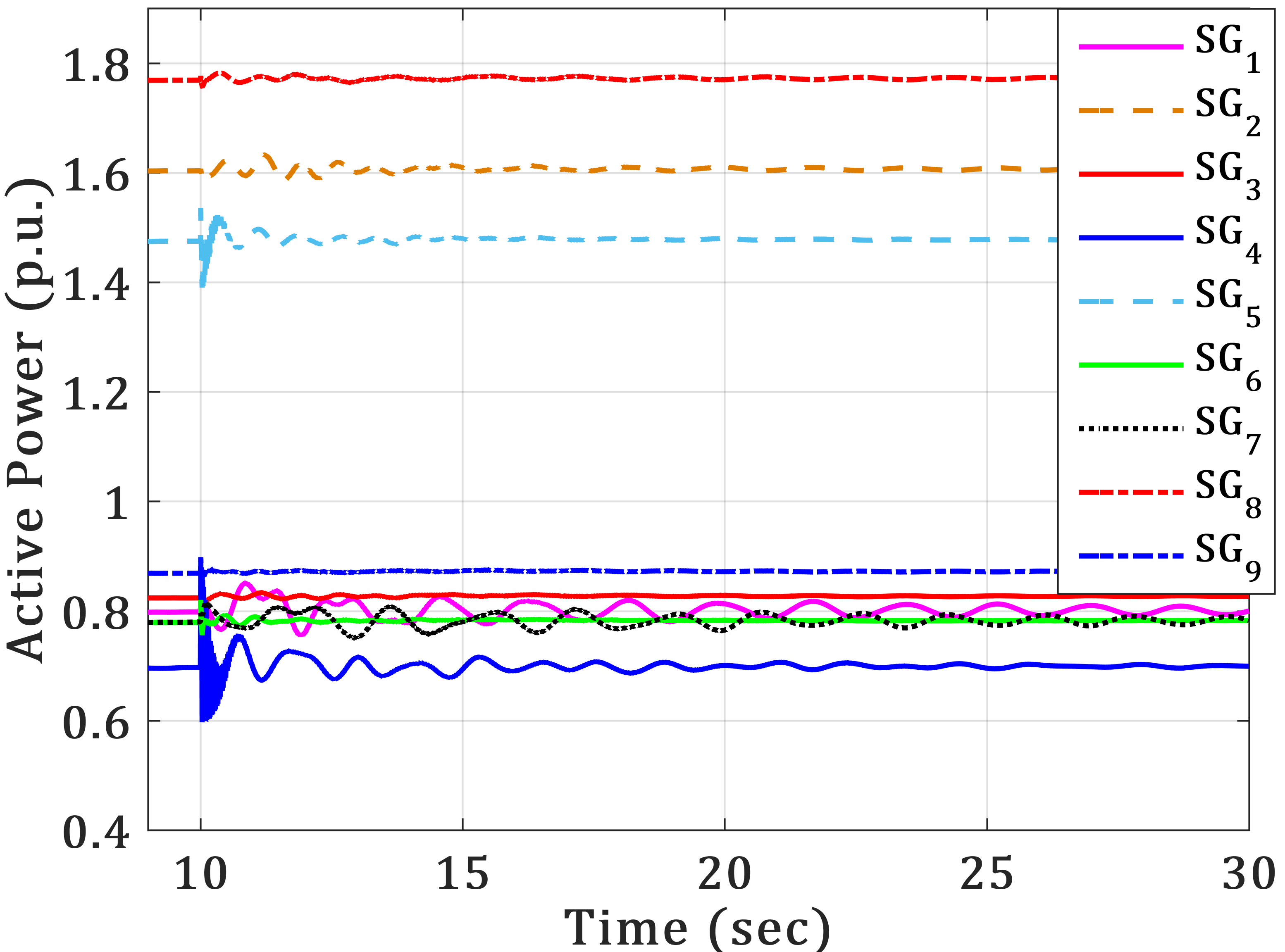}
         \vspace{-10pt}
         \caption{}
         \label{Plot_SG_P_gen_wo_volt_cont}
     \end{subfigure}
     \begin{subfigure}[h]{4cm}
         \centering
         \includegraphics[width=\textwidth]{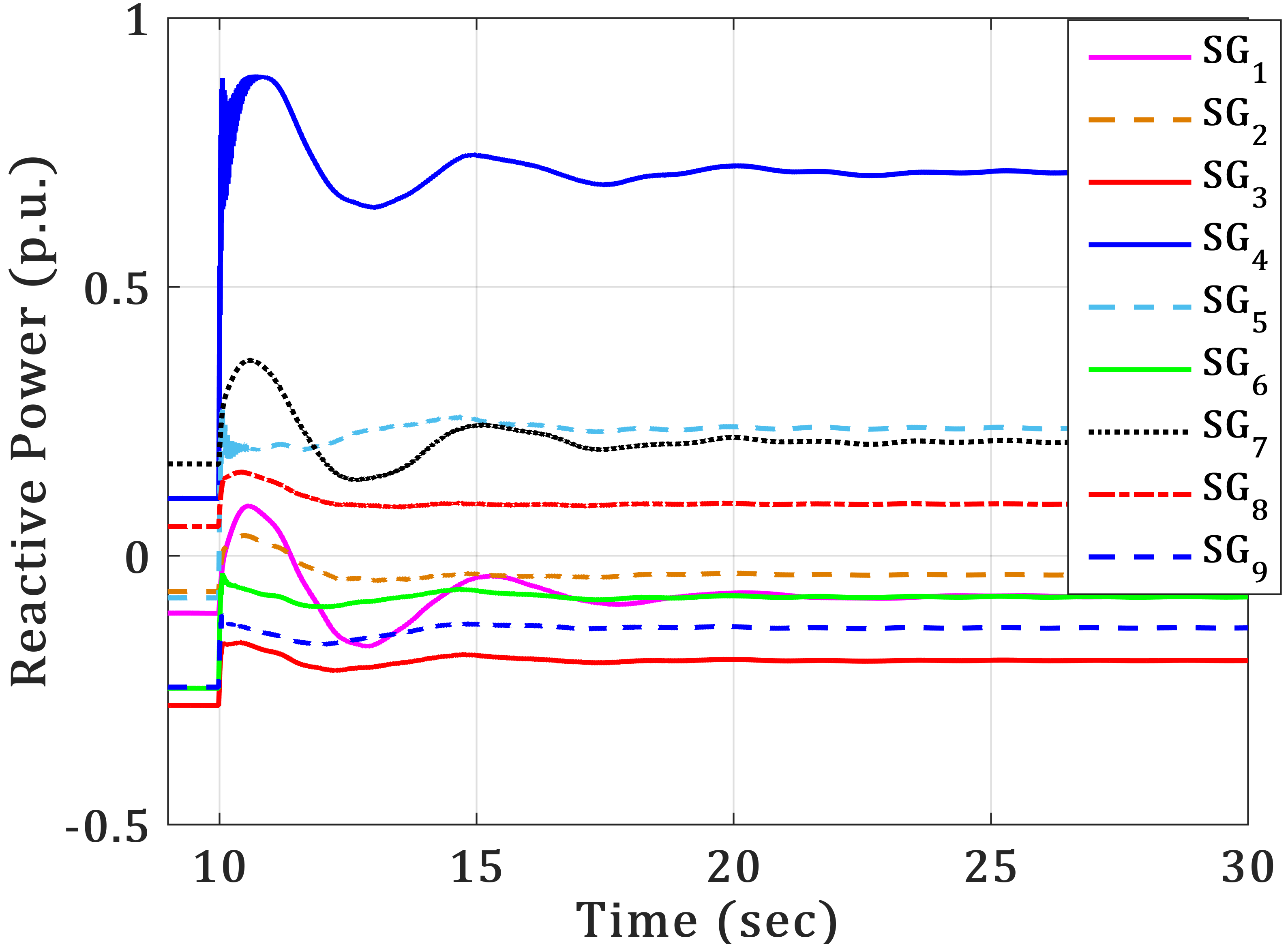}
         \vspace{-10pt}
         \caption{}
         \label{Plot_SG_Q_gen_wo_volt_cont}
     \end{subfigure}
          \vspace{-8pt}
        \caption{(a) Active power outputs of SGs, (b) Reactive power outputs of SGs with no hierarchical control: scenario (2)}
        \label{Plot_SG_PQ_gen_wo_volt_cont}
\end{figure}

Fig. \ref{Volt_Control_dynamics} also compares two different scenarios for this case: (1) Contingent bus 16 voltage with hierarchical control, (2) Contingent bus 16 voltage with no control applied, i.e., only SG-based primary control is active. Fig. \ref{Plot_SG_P_gen_wo_volt_cont} - \ref{Plot_SG_Q_gen_wo_volt_cont} show the active and reactive power output of SGs corresponding to the scenario (2). From Fig. \ref{Plot_SG_Q_gen_wo_volt_cont}, it is evident that the $SG_4$, being near to contingent bus 16, provides a significant reactive power support in the absence of proposed hierarchical control. However, in presence of hierarchical voltage control, the main reactive power support comes from $IBR_{class1}$ close to bus 16 before using the SG support. There is no significant active power contribution from SGs in scenario (2) as well. From Fig. \ref{Volt_Control_dynamics}, it is clear that Bus 16 voltage settles down near 0.9 p.u. in steady state for SG-based primary control. Our hierarchical control, in contrast, drives the contingent bus voltage closer to the safe band of [0.95, 1.05] p.u. and also shows better dynamic performance.\\
\vspace{-15pt}

\vspace{-10pt}

\subsection{Dynamic performance evaluation for simultaneous frequency and voltage control}

\begin{figure}[h]
     \centering
     \begin{subfigure}[h]{4cm}
         \centering
         \includegraphics[width=\textwidth]{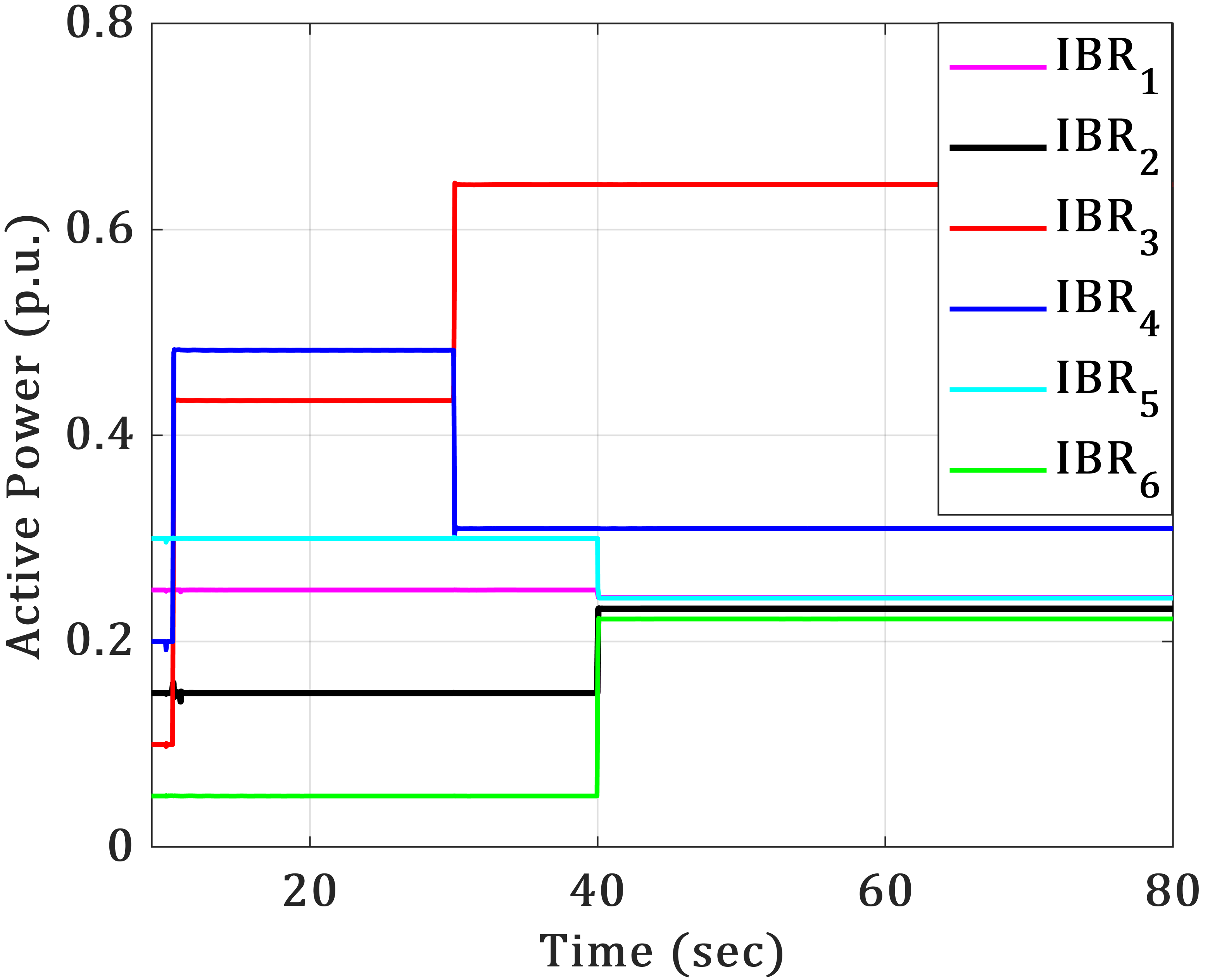}
         \vspace{-10pt}
         \caption{}
         \label{Plot_IBR_P_fV_cont}
     \end{subfigure}
     \begin{subfigure}[h]{4cm}
         \centering
         \includegraphics[width=\textwidth]{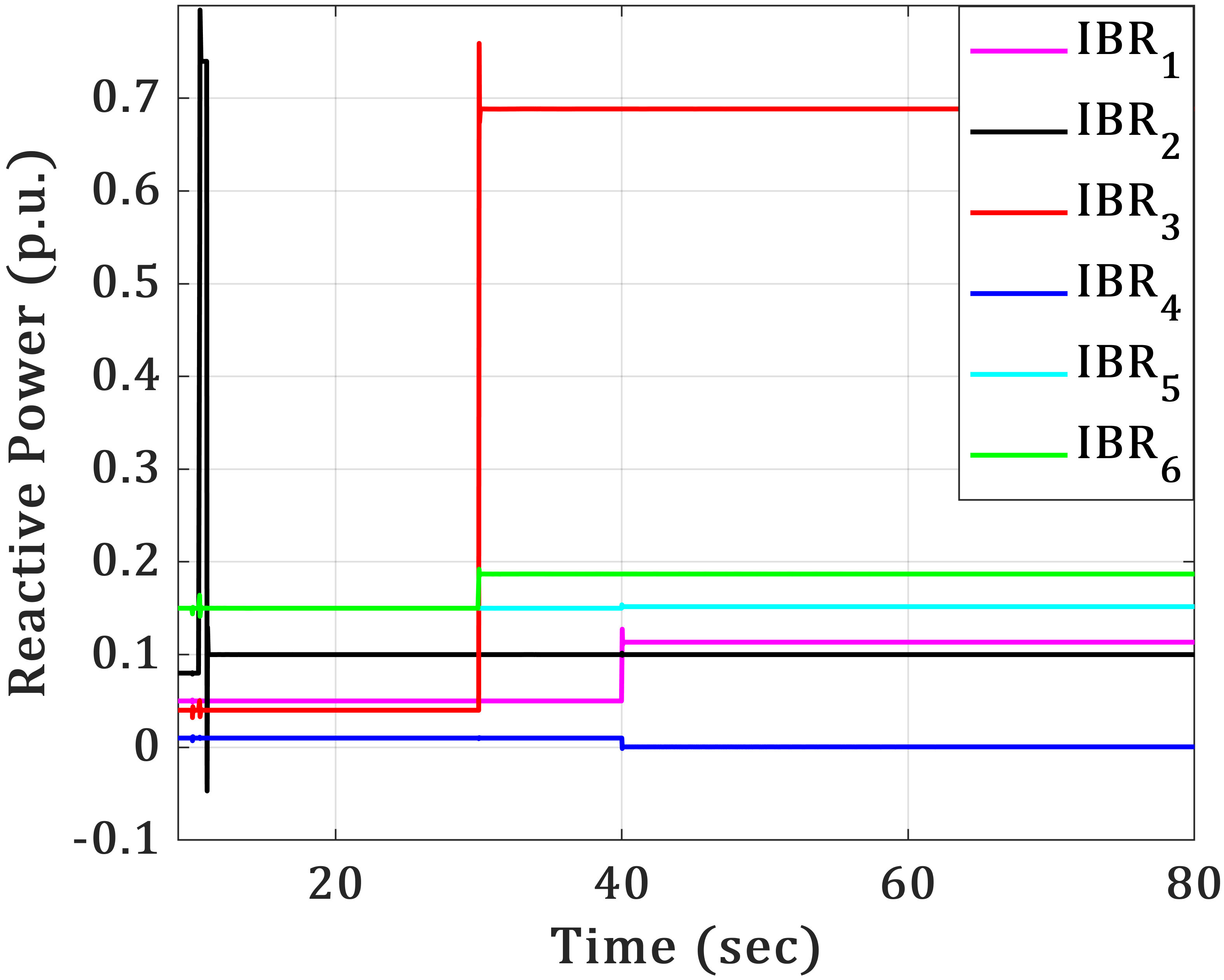}
         \vspace{-10pt}
         \caption{}
         \label{Plot_IBR_Q_fV_cont}
     \end{subfigure}
          \vspace{-8pt}
        \caption{(a) IBR active power outputs, (b) IBR reactive power outputs with simultaneous frequency and voltage control}
        \label{Plot_IBR_PQ_fV_cont}
\end{figure}

\begin{figure}[h]
     \centering
     \begin{subfigure}[h]{4cm}
         \centering
         \includegraphics[width=\textwidth]{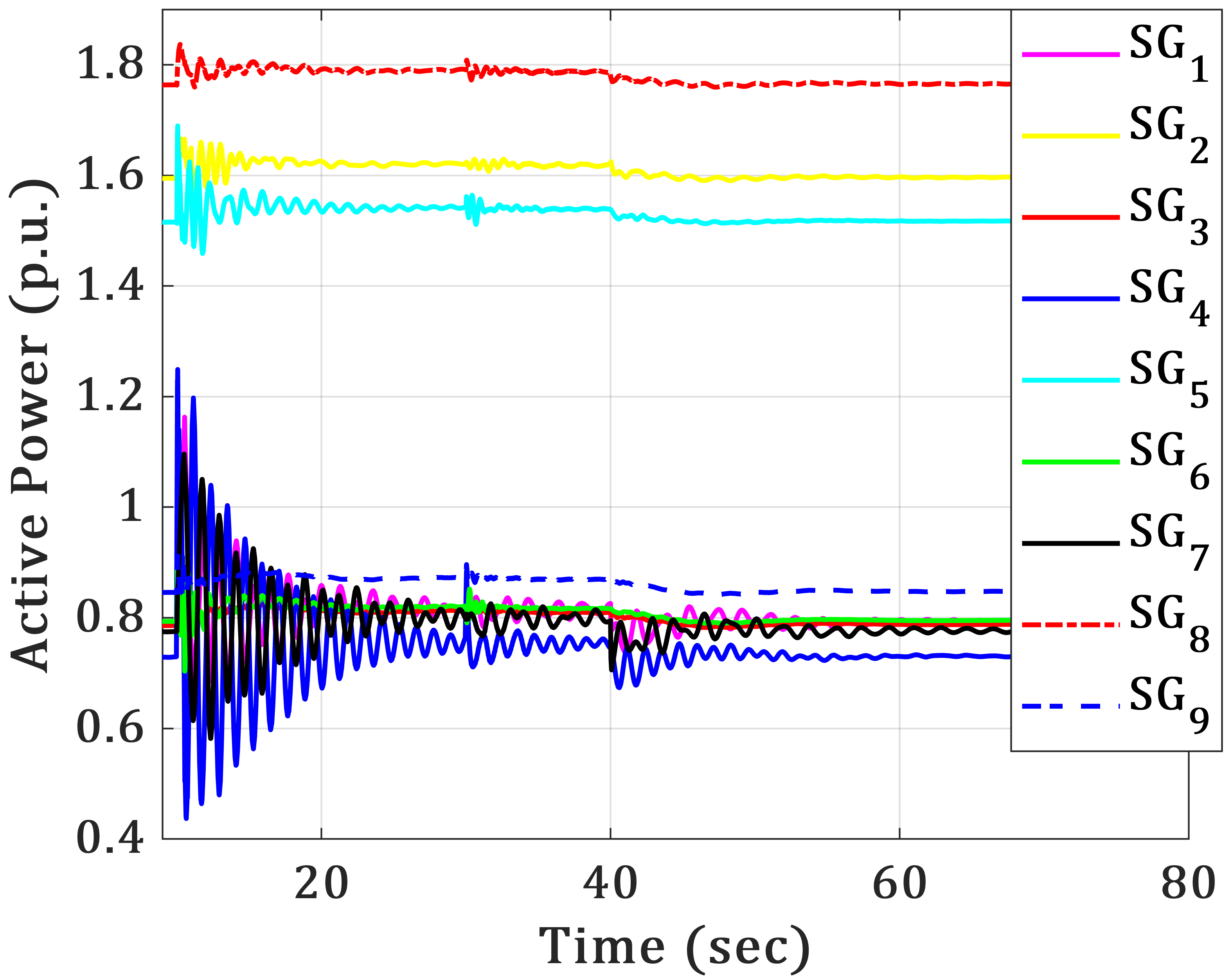}
         \vspace{-10pt}
         \caption{}
         \label{Plot_SG_P_fV_cont}
     \end{subfigure}
     \begin{subfigure}[h]{4cm}
         \centering
         \includegraphics[width=\textwidth]{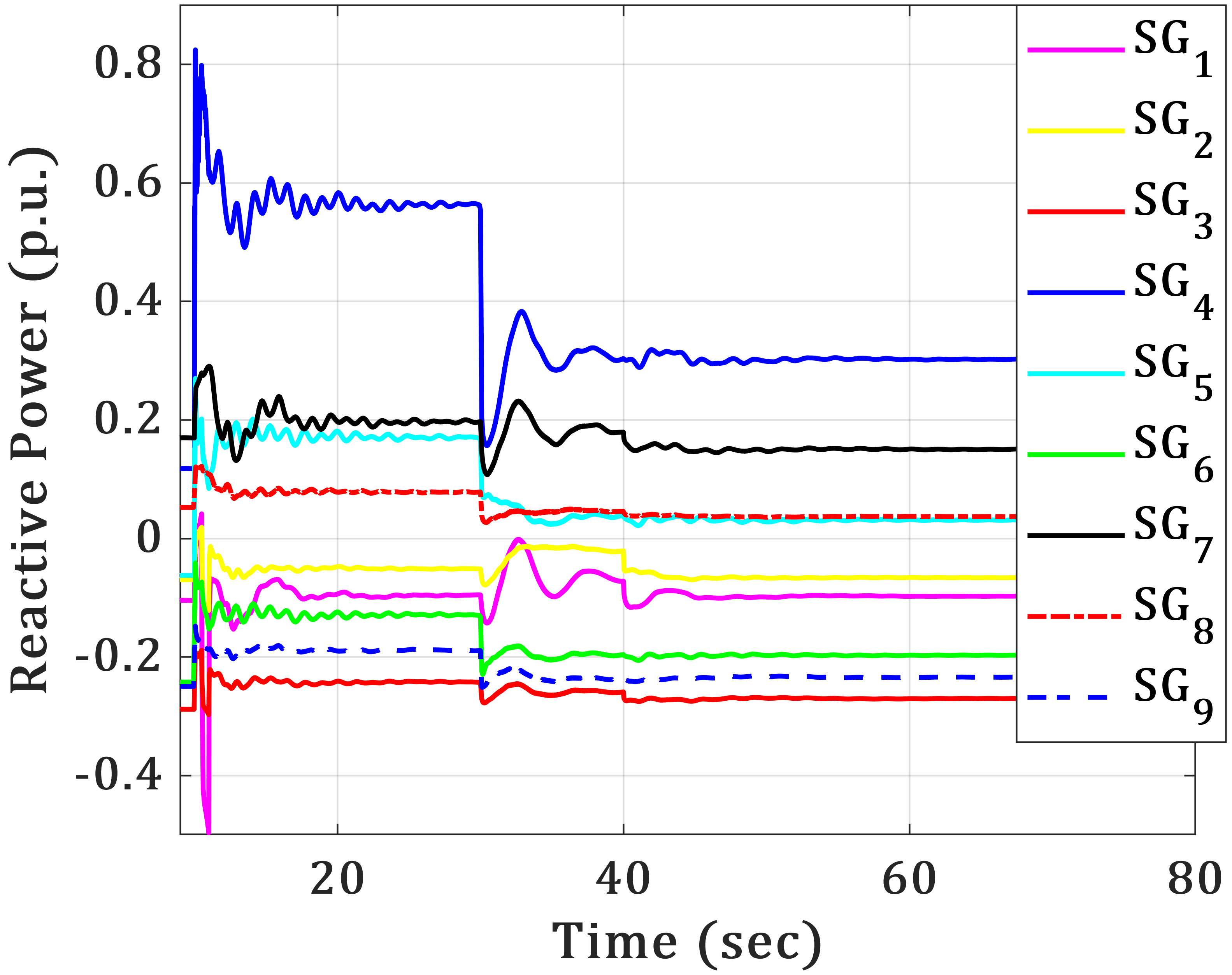}
         \vspace{-10pt}
         \caption{}
         \label{Plot_SG_Q_fV_cont}
     \end{subfigure}
     \vspace{-8pt}
        \caption{(a) SG active power outputs, (b) AG reactive power outputs with simultaneous frequency and voltage control}
        \label{Plot_SG_PQ_fV_cont}
\end{figure}

We simulate simultaneous change of active and reactive power of 80 MW and 50 MVAR respectively in the load connected to bus 16 at $t$ = 10 sec. This is 20.9\% and 43.4\% increase with respect to the total active and reactive power load respectively in the contingent area $A_1^{H_1}$. Even though contingent area has sufficient IBR headrooms to compensate for the aforementioned active power and reactive power change individually, the cumulative headroom is not adequate to make up for the concurrent change of real and reactive power. Therefore, simultaneous frequency and voltage control is applied following the steps shown in Fig. \ref{Freq_Volt_Flow_Chart} while prioritizing voltage control over frequency control. The reactive power setpoints of $IBR_{class1}$ and the active power setpoints of $IBR_{class2}$ are dispatched at $t$ = 10.5 sec, followed by the secondary voltage control setpoint actuation at $t$ = 11 sec. Please note that SG bus voltage setpoints are also changed at $t$ = 11 sec to involve the conventional reactive power resources in reactive power compensation. Finally, $IBR_{class2}$ are used in APPF-based secondary frequency control optimization to bring the balance of active and reactive power in the network, while also compensating for the extra network losses due to dispatched setpoints. The fast response of IBRs through active and reactive power setpoint dispatch is shown in Fig. \ref{Plot_IBR_P_fV_cont} - \ref{Plot_IBR_Q_fV_cont}. The  time response of the active and reactive power outputs of SG buses are represented in Fig. \ref{Plot_SG_P_fV_cont} - \ref{Plot_SG_Q_fV_cont}. Even though SGs participate in the reactive power compensation, their active power outputs settle down to pre-contingency outputs after the execution of simultaneous frequency and voltage control.\\
\vspace{-15pt}

\vspace{-18pt}

\begin{figure*}[h]
	\begin{center}
		\includegraphics[width=\textwidth]{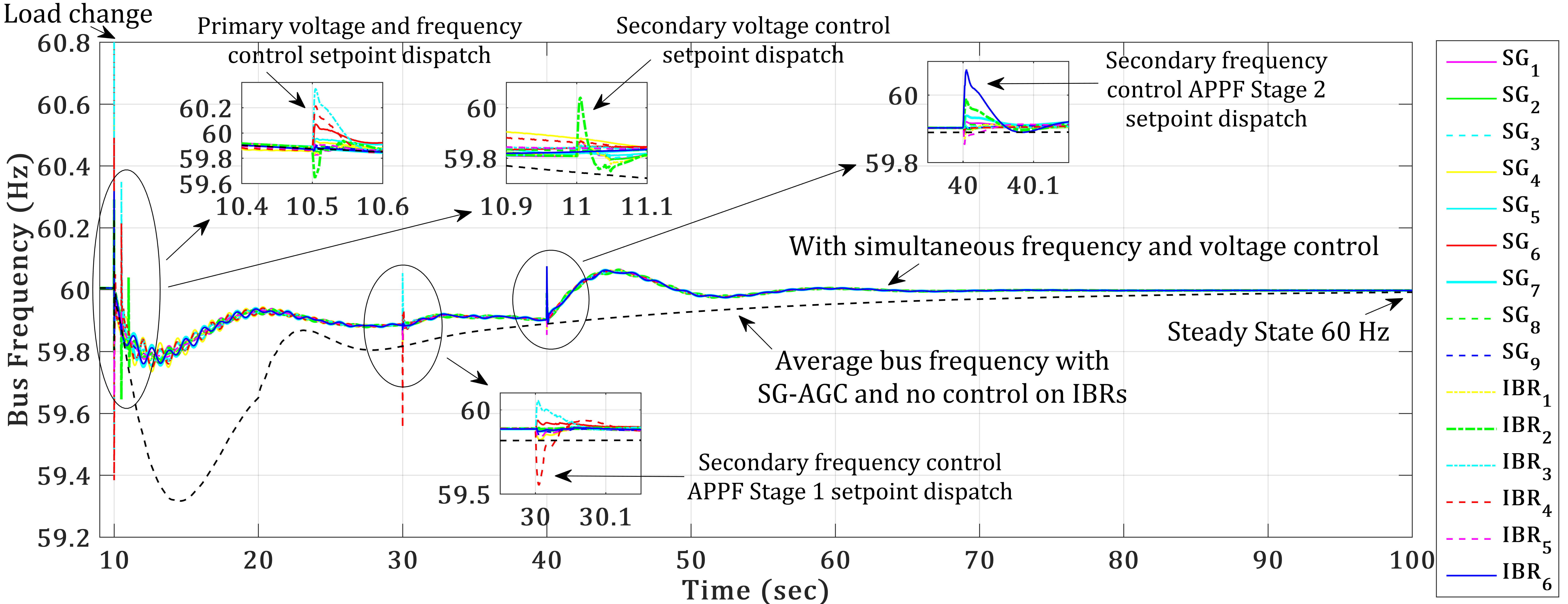}
		\caption{Comparison of frequency trajectories for APPF with simultaneous frequency and voltage control vs non-APPF}
		\label{Plot_freq_fV_cont}
	\end{center}
\end{figure*}

\begin{figure*}[h]
	\begin{center}
		\includegraphics[width=\textwidth]{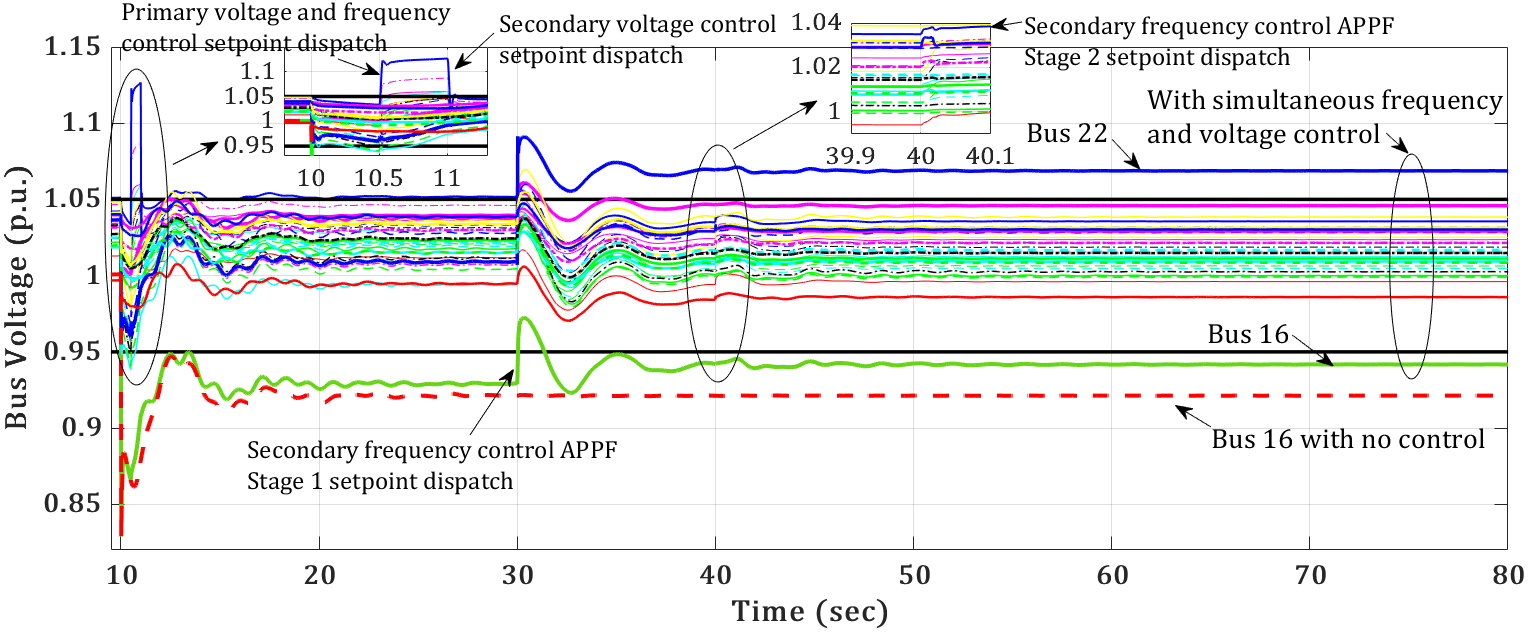}
		\caption{Comparison of voltage trajectories for APPF with simultaneous frequency and voltage control vs non-APPF}
		\label{Plot_volt_fV_cont}
	\end{center}
\end{figure*}

The  time  response of  the  bus  frequencies  are  shown  in  Fig. \ref{Plot_freq_fV_cont}. An  important observation is that both primary frequency control and APPF stage 1 affects the frequency trajectories due to the interplay between frequency and voltage control algorithm. In addition, both stages of APPF are involved due to the self-deficiency of the contingent area arising from simultaneous change of active and reactive power. In Fig. \ref{Plot_freq_fV_cont}, the average bus frequency trajectory for SG-AGC with no control on IBRs represents the scenario when only SG-based AGC control is active. Therefore, Fig. \ref{Plot_freq_fV_cont} shows that our  proposed  hierarchical  control  outperforms SG-based AGC control in terms of  dynamic  performance  of  the  frequencies  and  successfully drives the frequency to 60 Hz with lower settling time. The bus voltage trajectories are shown in  Fig. \ref{Plot_volt_fV_cont}. Similar to the hierarchical voltage control results of section-\ref{dynamic_volt_cont_sim}, most of the bus voltages stay within [0.95, 1.05] p.u., while voltage of bus 22 is pushed towards 10\% global bound to maximally utilize IBR$_4$ which is closest to the contingent bus. Fig. \ref{Plot_volt_fV_cont} also shows that simultaneous frequency and voltage control bring the contingent bus 16 voltage closer to the local band of [0.95, 1.05] p.u. compared to the case when no control is applied, i.e., only SG-based primary control is active.


\section{Conclusions}\label{conclusion}

We developed a hierarchical frequency and voltage control scheme for multi-area power systems with area-prioritized utilization of renewable energy resources. The idea of hierarchical control amounts to decomposing a centralized optimal control problem into a cascade of smaller dimensional optimization problems. Primary control is proposed based on fast re-dispatch of available IBR headroom in post-contingency condition. Secondary control is applied from an optimization based approach called APPF. The APPF methodology maximizes the usage of area-specific IBRs in post-contingency condition and minimizes the effect of disturbance in other areas with stable dynamics. On the other hand, IBRs act as fast actuators to improve the frequency nadir and transient voltage performance. In summary, the proposed control design improves dynamic frequency and voltage performance and also leads to the maximum utilization of renewable energy. One future direction of research would be to analyze how measurement noise, computation errors and other model uncertainties propagate from one stage of APPF to another, and how these uncertainties impact the accuracy of the overall solution.
\bibliographystyle{IEEEtran}
\bibliography{Reference}


\end{document}